\documentclass[manuscript,screen]{acmart}

\AtBeginDocument{%
  }

\setcopyright{acmlicensed}
\copyrightyear{2025}
\acmYear{2025}
\acmDOI{XXXXXXX.XXXXXXX}

\acmConference[n/a]{}{June 03--05,
  2025}{Woodstock, NY}
\acmISBN{978-1-4503-XXXX-X/18/06}

\usepackage{tikz}
\usepackage{booktabs}

\begin{document}

%%
%% The "title" command has an optional parameter,
%% allowing the author to define a "short title" to be used in page headers.
\title{Interrogating LLM design under a fair learning doctrine}

%%
%% The "author" command and its associated commands are used to define
%% the authors and their affiliations.
%% Of note is the shared affiliation of the first two authors, and the
%% "authornote" and "authornotemark" commands
%% used to denote shared contribution to the research.
\author{Johnny Tian-Zheng Wei}
\authornotemark[1]
\affiliation{%
  \institution{University of Southern California}
  \city{Los Angeles, CA}
  \country{USA}}
\email{jtwei@usc.edu}

\author{Maggie Wang}
\authornotemark[1]
\affiliation{%
  \institution{Princeton University}
  \city{Princeton, NJ}
  \country{USA}
}
\email{maggiewang@princeton.edu}

\author{Ameya Godbole}
\affiliation{%
  \institution{University of Southern California}
  \city{Los Angeles, CA}
  \country{USA}}
\email{ameyagod@usc.edu}

\author{Jonathan H. Choi}
\affiliation{%
  \institution{University of Southern California}
  \city{Los Angeles, CA}
  \country{USA}}
\email{jonchoi@law.usc.edu}

\author{Robin Jia}
\affiliation{%
  \institution{University of Southern California}
  \city{Los Angeles, CA}
  \country{USA}}
\email{robinjia@usc.edu}

%%
%% By default, the full list of authors will be used in the page
%% headers. Often, this list is too long, and will overlap
%% other information printed in the page headers. This command allows
%% the author to define a more concise list
%% of authors' names for this purpose.
\renewcommand{\shortauthors}{Wei and Wang et al.}

%%
%% The abstract is a short summary of the work to be presented in the
%% article.
\begin{abstract}
The current discourse on large language models (LLMs) and copyright largely takes a ``behavioral'' perspective, focusing on model outputs and evaluating whether they are substantially similar to training data. However, substantial similarity is difficult to define algorithmically and a narrow focus on model outputs is insufficient to address all copyright risks. In this interdisciplinary work, we take a complementary ``structural'' perspective and shift our focus to how LLMs are trained. We operationalize a notion of ``fair learning'' by measuring whether any training decision substantially affected the model's memorization. As a case study, we deconstruct Pythia, an open-source LLM, and demonstrate the use of causal and correlational analyses to make factual determinations about Pythia's training decisions. By proposing a legal standard for fair learning and connecting memorization analyses to this standard, we identify how judges may advance the goals of copyright law through adjudication. Finally, we discuss how a fair learning standard might evolve to enhance its clarity by becoming more rule-like and incorporating external technical guidelines.
\end{abstract}

%%
%% The code below is generated by the tool at http://dl.acm.org/ccs.cfm.
%% Please copy and paste the code instead of the example below.
%%


%%
%% Keywords. The author(s) should pick words that accurately describe
%% the work being presented. Separate the keywords with commas.
\keywords{Copyright, LLMs, regression}
%% A "teaser" image appears between the author and affiliation
%% information and the body of the document, and typically spans the
%% page.
%\begin{teaserfigure}
%  \includegraphics[width=\textwidth]{sampleteaser}
%  \caption{Seattle Mariners at Spring Training, 2010.}
%  \Description{Enjoying the baseball game from the third-base
%  seats. Ichiro Suzuki preparing to bat.}
%  \label{fig:teaser}
%\end{teaserfigure}

\received{20 February 2007}
\received[revised]{12 March 2009}
\received[accepted]{5 June 2009}

%%
%% This command processes the author and affiliation and title
%% information and builds the first part of the formatted document.
\maketitle

\section{Introduction}

% large language models pose issues for copyright
% these issues are beginning to be litigated
% 

% backdrop of what is being litigated
Large language models (LLMs) present new challenges to copyright law \cite{Franceschelli_Musolesi_2022, henderson_2023_foundation}. While LLMs hold transformative potential in many domains \cite{Bommasani2021FoundationModels,Franceschelli2024}, these models are trained on vast amounts of textual data which often includes copyrighted material \cite{chang-etal-2023-speak}. Technical studies have shown that LLMs can reproduce parts of their training data \cite{Carlini2020ExtractingTD}, although under normal use, their outputs are typically novel \cite{merrill-etal-2024-evaluating}. In the U.S., it is unclear whether training these models on copyrighted material violates the authors' exclusive copyright of that material. As LLM outputs fall in between creativity and copying, existing legal categories break \cite{lee2024talkinboutaigeneration}.

The current discourse on LLMs and copyright largely takes a ``behavioral'' perspective, focusing on model outputs and evaluating whether they are substantially similar to training data \cite[][inter alia]{sobel2017fairuse, DBLP:conf/cvpr/SomepalliSGGG23, ippolito-etal-2023-preventing, scheffler_2022_formalizing}. However, substantial similarity is difficult to define algorithmically, and similarity alone is a narrow view of copyright. Complementary to the behavioral perspective, a ``structural'' perspective would focus on how the model was trained. \textbf{Our work explores a structural perspective and theorizes a fair learning doctrine to clarify acceptable LLM design.} To this end, we conduct an interdisciplinary exploration of judicial decision-making for LLMs. 
%We identify why training decisions matter under relevant copyright law doctrines and demonstrate causal and correlational methods to evaluate the effects of those decisions on memorization. Based on the technical methods, we propose a fair learning doctrine for courts to apply in evaluating LLM design choices. 
Our work consists of three parts:
\begin{itemize}
    \item In \S\ref{sec:legal_review}, we review U.S. copyright law and emphasize that the legal determination of fair use can turn on technological design decisions. Here, a ``behavioral'' perspective alone is an incomplete view of copyright, and this work takes a complementary ``structural'' perspective by evaluating LLM design decisions, which presents an opportunity to develop a notion of ``fair learning.''

    \item In \S\ref{sec:deconstruct}, we operationalize a notion of ``fair learning'' by measuring whether a model's training decisions substantially affected its memorization, and examine the training of Pythia, an open-source LLM \cite{DBLP:conf/icml/BidermanSABOHKP23}. Since the discretion in training Pythia was in creating its training dataset (22 datasets were combined and high quality data upweighted), we demonstrate two analyses: a stronger causal analysis on the effects of upweighting, and a weaker correlational analysis to simulate dataset ablations and study dataset curation. We make a few factual determinations about Pythia's training decisions and discuss how to interpret different types of statistical evidence.

    \item In \S\ref{sec:formalizing}, we propose a fair learning doctrine to clarify acceptable LLM design. By shifting the legal focus to design choices, we identify a number of desirable normative outcomes which the judicial system can accomplish through adjudication. We end on a discussion of how fair learning may evolve to enhance its clarity and predictability by becoming more rule-like and incorporating external technical guidelines.
\end{itemize}

\section{Review of U.S. copyright law} \label{sec:legal_review}

U.S. copyright law is rooted in the Constitution, which grants Congress the power to promote the progress of science and the useful arts \cite{us_const_art1_sec8_cl8}. To accomplish this, copyright law seeks to balance incentivizing creativity and ensuring public access to creative works. Copyright law incentivizes creativity by granting authors exclusive rights over reproduction, distribution, and the creation of derivatives of their works \cite{usc_17_106}. To ensure public access, copyright limits those exclusive rights through doctrines like fair use, which provides exceptions for the use of copyrighted materials \cite{usc_17_107}. Historically, granting fair use was understood to benefit the public by allowing new uses of copyrighted works \cite{sobel2017fairuse}. 

Private corporations training LLMs on copyrighted material argue that their training constitutes fair use, excusing them from liability for copyright violations. Allowing fair use in this context would allow private corporations to benefit from publicly available, copyrighted material. We do not take normative positions on copyright (and we acknowledge the opportunity to consider whether copyright law continues to be adequate \cite{CrootofArd2021} and its radical reimagination \cite{lemley2023generativeAI}). 
Courts will need to make their own decisions in balancing innovation against public interests, and our work explores judicial decision-making for LLMs to better enable courts to advance copyright's existing normative goals. 

\subsection{Substantial similarity and fair use}

In a series of high-profile lawsuits, courts are grappling with how to adapt traditional copyright law principles to LLMs \cite{masterlist_ai_lawsuits_2024, atkinson2024legalrisktaxonomygenerative}. This section reviews copyright law, emphasizing the role of technology design. When doctrinal considerations intersect with technological design, legal outcomes can turn on how a technology operates or is structured.

\paragraph{Substantial similarity} Primary copyright infringement occurs when a protected work is used without authorization in a way that violates the author’s exclusive rights. To establish infringement, courts first assess whether the defendant had access to the copyrighted work and whether the work in question is substantially similar to the protected material.\footnote{E.g., \citeauthor{arnstein1946} (discussing the “substantial similarity” test and the importance of both access and similarity in determining infringement).} In \citeauthor{feist1991}, the Supreme Court clarified that while facts are not copyrightable, their expressive elements, such as creative selection or arrangement, can be. A substantial similarity analysis must therefore be constrained to the protected, expressive elements of a work. 

Tests for substantial similarity generally attempt to separate expressive elements, which are protected by copyright, from non-expressive elements that are not eligible for protection, such as ideas, facts, and functional aspects. Only the expressive elements are then compared to determine if the works are substantially similar. For instance, \citeauthor{computerassociates1992} introduced the "abstraction-filtration-comparison" test for code, which attempts to break down code into structural components, filter out the non-expressive elements, and compare the remaining expressions for substantial similarity. The boundary between factual and expressive elements can vary across tests, and can also shift across mediums, as the expressive elements in code may differ from those in literary works or the visual arts \cite{lim2021substantial}.

\paragraph{Fair use} To ensure public access to copyrighted works, limited use of copyrighted works without authorization is exempt under fair use \cite{usc_17_107}, which is an affirmative defense that legitimizes conduct that would otherwise be copyright infringement. A determination of fair use is based on four factors: (1) The purpose and character of the use, (2) the nature of the copyrighted work, (3) the amount and substantiality of the portion used, and (4) the effect of the use on the market for the original. \citeauthor{campbell1994} illustrates the application of these factors, where the Supreme Court ruled that a parody was fair use, emphasizing the transformative nature of the parody and the minimal harm to the market for the original work.

Modern copyright cases often hinge on fair use. In \citeauthor{field2006}, the technology in question was Google’s caching feature, which stored web pages for improved search performance. Since the cache contains exact copies of websites, the copies were substantially similar. However, the court ruled in favor of Google, recognizing that Google’s use of robots.txt was industry standard for respecting opt-out preferences. Similarly, in \citeauthor{perfect10_2007}, Google's design choice to use small, low-resolution thumbnails rather than full-size images was key to the fair use analysis, as it minimized market harm and facilitated image search and discovery rather than content consumption. Both cases highlight how design decisions can influence judicial decisions on what is fair, and that courts can evaluate whether the design reflects a reasonable balance between the rights of copyright holders and the broader public interest in enabling new functionalities.

\paragraph{Fair use and LLMs} Fair use is especially important to contemporary debates over LLMs and copyright because all parties generally agree that companies training LLMs prime facie violate copyright law by creating and storing copies of copyrighted materials in internal databases for use in training models; the key debate, then, is whether this practice constitutes fair use, which is why companies like OpenAI emphasize fair use in their legal arguments \cite{openai_journalism}.

Importantly, the ``purpose and character'' prong of fair use implicates the motives and the good conduct of a potential copyright violator. The Supreme Court held in \citeauthor{harper1985} that ``the propriety of the defendant's conduct'' is relevant to the ``character'' of the use, noting that ``[f]air use presupposes `good faith' and `fair dealing'.'' Thus, when companies training LLMs claim the defense of fair use, courts should evaluate whether they made appropriate training decisions and took reasonable measures to prevent reproduction of copyrighted materials. 

\subsection{Behavioral and structural perspectives}

Most lawsuits about copyright violation in LLM training involve a plaintiff, who claims substantial similarity, and a defendant, who asserts fair use. Behavioralism and structuralism are then two complementary approaches to address copyright for LLMs (terminology due to \citet{greenberg2016technology}). Here we selectively emphasize the advantages of the former and disadvantages of the latter to open up our exploration of structuralism.

\paragraph{Behavioralism} The first approach focuses on the behavior, or outputs of a generative model, to ensure that they are never substantially similar to any copyrighted work. Much of the machine learning literature on copyright adopts this behavioral perspective, measuring the similarity of model outputs to pieces of the training data \cite[][inter alia]{DBLP:conf/cvpr/SomepalliSGGG23, merrill-etal-2024-evaluating, freeman2024exploring}. Extending these measurements, some works study technical mitigation to reduce this output similarity \cite{wei2024evaluating, chen-etal-2024-copybench, ippolito-etal-2023-preventing}. Preventing output similarity has analogs in differential privacy, which can prevent training data reproduction with statistical guarantees \cite{pmlr-v202-vyas23b} and machine unlearning, which can erase copyrighted data after training \cite{meng_locating_2022}. As outlined in \citet{wei2024evaluating}, combining mitigation techniques with a safe harbor provision, such as a notice-and-takedown regime modeled after the DMCA, provides a sociotechnical solution to addressing copyright issues for generative models.

However, mitigation methods require an algorithmic formulation of substantial similarity. Copyright scholars have long lamented the inconsistency of substantial similarity \cite{lim2021substantial}, as judicial interpretation of what elements are expressive is not uniform and tests for substantial similarity are not consistently applied. This subjectivity may resist technical operationalization  \cite{cooper2024machineunlearningdoesntthink}. Without a precise algorithmic determination of copyright's substantial similarity, problematic outputs risk falling through. On the other hand, encouraging the application of mitigation techniques risks overregulating model outputs beyond the intent of the law \cite{burk2005legal}. 

Ultimately, behavioral analysis alone is neither necessary nor sufficient to establish copyright violation. It is not necessary because models could still infringe even if their outputs are never similar to copyrighted materials, so long as copying of copyrighted materials occurred when preprocessing training data (which is universally agreed) or potentially if the models themselves implicitly contain copies of copyrighted materials. Behavioral analysis is also not sufficient because transformer-based language models are expressive generation models which assign non-zero probability to every piece of text \cite{slp_book}. This makes them susceptible to adversarial prompting \cite{wallace-etal-2019-universal, schwarzschild2024rethinkingllmmemorizationlens}, and with the right prompt, they can generate any piece of text; thus the ability to extract copyrighted text from a LLM cannot be taken as per se copyright infringement on its own, to which OpenAI has claimed in their lawsuits \cite{brittain2024}.  The key legal question is not whether the outputs are substantially similar, but whether the fair use exception applies.

\paragraph{Structuralism} The second approach focuses on the structure, or design, of a generative model, to ensure that the model is learning and memorizing only non-expressive facts and ideas from its training data. Normatively, it may be desirable for model developers to deploy models that are copyright safe by design \cite{elkin-koren2017fairuse}, which requires developers to make careful decisions during training. These decisions can include which datasets are selected, how they were filtered, whether they were upweighted, and also the use of memorization-reduction techniques such as differential privacy \cite{carlini2019}. The study of memorization is an active research area \cite{hartmann2023sokmemorizationgeneralpurposelarge} which has begun to understand the effects of basic design decisions on memorization \cite{tirumala-memorization-2022, pmlr-v162-kandpal22a}. These design decisions matter, as generative models can differ significantly in their capacity to memorize training data, presenting varying levels of copyright risk \cite{sag2023copyright, lee2024talkinboutaigeneration}. Training decisions can be systematically evaluated and optimized, and the machine learning literature presents many examples of rigorously studying LLM design decisions for the purposes of improving performance \cite{lee-etal-2022-deduplicating, longpre-etal-2024-pretrainers}. How then could we formalize a notion of ``fair learning'' \cite{lemley2020fair}? This is the heart of our inquiry, where we provide an operationalization of ``fair learning'' and relate it to judicial adjudication.

\section{Deconstructing an open-source LLM} \label{sec:deconstruct} 

The design choices of an LLM can intersect with doctrinal considerations in copyright law. For instance, in \citeauthor{nytimes_v_microsoft}, the NYTimes alleges that Microsoft not only used but upsampled their data during training. This alleged upweighting raises concerns about the Amount and Substantiality of the Portion Used (factor three of fair use), as such a training decision increases the risk of memorization of the upweighted data. Treating machine learning as a black box can interfere with judicial assessment and hinder a complete doctrinal analysis, where \citet{selbst2024deconstructing} calls for courts to deconstruct design decisions. Our investigation here is a direct response to this call. In this section we put Pythia \cite{DBLP:conf/icml/BidermanSABOHKP23}, an open-source LLM, on trial and ask: was the pretraining of Pythia ``fair''? While we uncover several technical facts about Pythia, our goal is not to make generalizations about LLMs. Each LLM would require a separate analysis, but we illuminate the deconstruction process to open up a legal discussion in \S\ref{sec:formalizing}.

\subsection{A case study on Pythia}
% In this section we put the Pythia model family \cite{DBLP:conf/icml/BidermanSABOHKP23} on trial and ask: was the pretraining of Pythia 'fair'? While Pythia differs from commercial LLMs like ChatGPT in scale and capability, we believe the analysis of its design decisions around data curation and weighting are illustrative. of choices that courts will need to evaluate. 

We examine the discretion in training Pythia \cite{DBLP:conf/icml/BidermanSABOHKP23}, an open source LLM. Its open-source nature allows us to apply a wide range of post-hoc analyses. Pythia employs a standard transformer architecture \cite{transformers}, and the model developer's design decisions mainly centered on creating the training data. The training data was created in two steps:
\begin{itemize}
    \item \textit{Curation}: Pythia models are trained on the Pile, where \citet{pile} selected “22 diverse and high-quality datasets, including both established natural language processing datasets and several newly introduced ones”. Deciding which data to include or exclude can implicitly reflect the value of the model developers  \cite{gururangan-etal-2022-whose}.
    
    \item \textit{Upweighting}: \citet{pile} choose to upweight datasets through duplication and “increase the weights of higher quality components, with certain high-quality datasets such as Wikipedia being seen up to 3 times”. These upweighting decisions explicitly encode the values of the model developers \cite{birhane_2022_values}.
\end{itemize}
Pythia's pretraining methodology is not representative of all LLMs, and the purpose of our case study is to demonstrate the statistical analysis of training decisions rather than producing empirical findings generalizable to all LLMs. Decisions that explicitly and implicitly upweight data are also present in Llama 3, a state-of-the-art LLM \cite{grattafiori2024llama3herdmodels}. For instance, the model was trained using simulated annealing, where certain high quality data was explicitly emphasized during the later stages of training. The model's training data was also curated with a quality classifier, which may implicitly upweight certain text. Some of the decisions of Llama 3 will be analyzable with the methods presented here, and we want to emphasize that the effects of training decisions on memorization are amenable to analysis.

\subsection{Operationalizing ``fair learning''} \label{sec:operationalizing}

To understand whether the training of Pythia was ``fair'', we operationalize ``fair'' by measuring whether any training decision substantially affected the model's memorization. We take inspiration from doctrines such as material contribution for contributory infringement and design defects for product liability to assess whether specific choices were substantial factors in causing harm. 

\paragraph{Soundness} Making sound measurements will be important to the assessment of training decisions. Ideally, we would compare a model's memorization against other models trained with different training decisions. However, training even one LLM is difficult and an ideal measurement is not always practical. The next best thing is to conduct a causal analysis, measuring the effects of each training decision while controlling for confounders. A randomized control trial is the most straightforward way to conduct a causal analysis and the randomization in the train/test split allows us to causally analyze the decision to upweight in \S\ref{sec:upweighting}.  However, some design choices may not be amenable to causal analysis, due to a lack of randomization or non-linear and diffuse effects. In such cases, we may only be able to derive correlational results. While these methods lack the rigor of causal inference, they can still offer valuable insights in patterns of model behavior which can be followed up with stronger experimental observations. We conduct a correlational analysis in \S\ref{sec:curation} and observe how dataset overlaps relate to model memorization.

\paragraph{Memorization metrics} To quantitatively assess model memorization, we choose four complementary metrics that capture different aspects of how models memorize and reproduce content:
\begin{itemize}
    \item \textit{Loss.} We measure model loss on text, which directly relates to the model's likelihood of generating that text. A lower loss indicates stronger memorization, and loss is commonly used in privacy research \cite{hartmann2023sokmemorizationgeneralpurposelarge}. For example, model loss is often used as a baseline in membership inference and can reveal whether a specific datapoint was part of the training set due to overfitting \cite{DBLP:conf/sp/ShokriSSS17}. While \citet{elkinkoren2024copyrightreducedprivacy} point out that technical notions of privacy can diverge with copyright, it is a useful starting point for operationalizing memorization concerns in the context of copyright. Since loss reflects model memorization, it more closely relates to the question of whether LLMs are derivative works of their training data.

    \item \textit{MinK\%.} \citet{shi2024detecting} propose MinK\%, a membership inference metric which averages only the loss of the K\% lowest probability tokens (instead of all tokens as is done for loss). As a membership inference technique, this metric seeks to isolate memorization signal from the loss, based on the intuition that seen text is less surprising to an LLM, whereas unseen text is more likely to contain surprising words.

    \item \textit{Token accuracy.} We measure the token-wise accuracy of a model on 10-token snippets of text, when prompted with the preceding 40 tokens. This metric closely aligns with memorization metrics used in training data extraction attacks \cite{carlini_extracting_2021}. Unlike loss, which reflects memorization indirectly, token accuracy is directly measured on model outputs.  Since token accuracy reflects an output similarity, it more directly relates to whether LLM outputs are substantially similar to their training data.

    \item \textit{Verbatim memorization.} This metric measures whether the model can perfectly predict the 10-token continuation of text conditioned on the preceding 40 tokens (i.e. token accuracy is 100\%) which follows the $k$-extractable memorization definition in \citet{carlini_extracting_2021}.
\end{itemize}
Determining whether a model infringes on copyrights is a complex judicial decision, and our measurements of memorization are not intended to automate that legal judgment. Rather, our goal is to establish a quantitative framework for analyzing design choices in model training. Similar to \citet{scheffler_2022_formalizing}, the presentation of quantitative results can form one part of a larger legal assessment, and we further elaborate on in \S\ref{sec:formalizing}.

\paragraph{Measuring memorization} Metrics like loss are used as both performance metrics \cite{kaplan2020scalinglawsneurallanguage} and memorization metrics, and theoretically, a model's generalization ability is closely tied to its ability to memorize \cite{vitaly_2020_does}. For copyright, only memorization of expressive elements are important, but these metrics equally apply to all kinds of data regardless of the nature of the data. Our memorization metrics measure more than the memorization of expressive elements, and are upper bounds which are sufficient for the purposes of minimization. To disentangle both generalization from memorization and expressive from non-expressive elements is extremely difficult, and better metrics of substantial similarity may complement our operationalization here \cite{hacohen2024similaritiescreatedequalleveraging, scheffler_2022_formalizing}. % We choose these metrics as they are technically sound and related to copyright concerns, and emphasize that our analysis framework is agnostic to the metric applied.

\subsection{A causal analysis at the document level} \label{sec:upweighting}

We first analyze the decision to upweight and whether upweighting a document substantially affected the document's memorization. 
%Since prior research establishes a strong connection between the number of occurrences of a document in the training data and its propensity for being memorized, 
%A plaintiff could reasonably allege that the decision to upweight any one of their documents contributes substantially to the LLM's memorization of that document.
%While most plaintiffs are not concerned with the memorization of single documents, 
This setting allows us to demonstrate a strong causal analysis that can be used to analyze a range of LLM training decisions. We frame our investigation through a counterfactual question: how would the memorization of a document change if it wasn't upweighted? This question is well suited to causal analysis due to the randomized train/test split of the Pile.

\paragraph{Train/test split} The randomization present in the train/test partition allows us to conduct a randomized control trial (RCT) to measure the effects of dataset upweighting. For instance, some documents in Wikipedia were randomly assigned to the training set and received a weight of three, while others were assigned to the test set and effectively received a weight of zero. With this randomization, any factors confounding Wikipedia memorization across documents are evenly spread across the training and test sets, so their memorization difference provides an unbiased estimate of upweighting's causal effect. Using random assignment to control for confounders is conceptually simple and compares a control group against a treatment group \cite{angrist_2009_mostly}.

\begin{figure}[t]
\centering
\includegraphics[]{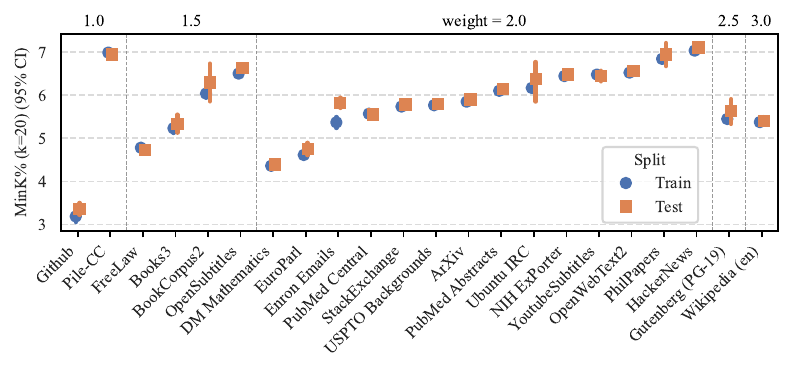} % denotes a regression
\caption{MinK\% (where K=20) for the Pythia 6.9b model across training (blue) and test (orange) sets, grouped by dataset component. MinK\% scores are calculated and averaged on up to 1k documents for the first 256 tokens (lower indicates better memorization). Error bars denote 95\% CI. The difference in the metric between train and test represents the causal effect of upweighting a single document. Differences are grouped by upweight value, assigned by the creators of the Pile. Differences for other metrics are presented in Figure \ref{fig:all_metrics}. All metrics are similar across the train and test splits indicating that upweighting a single document has little effect on memorization.}
\label{fig:plot7bmink}
\end{figure}
% Old Caption: Mean cross-entropy loss (with 95\% CI error bars) for the Pythia-6.9b model across training (blue) and test (orange) sets, grouped by data source. Vertical lines mark boundaries between upweight groups (upweight labeled along the top). Within upweight boundaries, sources are ordered by their train-loss average values. The x-axis labels the data sources, while the y-axis shows the average loss (lower indicates better memorization). Notably, the similar train and test losses indicate little discrepancy in memorization across data sources, reflecting the negligible affect of explicit upweighting on memorization.
\paragraph{Assumptions of an RCT} RCTs are conceptually simple but rely on key assumptions to correctly estimate the causal effect of an intervention. Our experimental design based on the randomization in the train/test partition already satisfies several key causal inference requirements. To satisfy the Stable Unit Treatment Value Assumption (SUTVA), a prerequisite for valid causal inference \cite{ding2023coursecausalinference}, we further require that there is no interference and that there are no spillover effects. Although the randomized binary treatment makes upweighting an ideal case for demonstrating causal analysis of LLM design decisions, important limitations remain. Specifically, the presence of one document in training could spillover to how other documents are memorized, thereby violating SUTVA. Even though SUTVA is violated, the effect of an individual datapoint on the model is generally small, so it may be an appropriate approximation.

\paragraph{Regression} Here we develop basic notation for a regression and then provide its causal interpretation. We skip a detailed discussion of the potential outcomes framework, which may be found in \citet{angrist_2009_mostly} or \citet{ding2023coursecausalinference}.
We model the relationship between model memorization and the upweighting as:
\begin{equation}\label{eq:explicit_weights}
    Y_i = \alpha + \beta_1 D_i + \epsilon_i
\end{equation} where $Y_i$ is the $i$-th document's memorization (as measured by the metrics in \S\ref{sec:operationalizing}), $D_i$ is a dummy variable set to 1 if the $i$-th document was in the training set, 0 otherwise, and $\epsilon_i$ is an error term for the unaccounted loss.
In this regression $\alpha$ represents the intrinsic loss of the model on this dataset, and $\beta_1$ represents the effect of upweighting the data (each document may be upweighted up to 3 times). If all the assumptions for causal analysis are met, the coefficients of the regression will have a causal interpretation \cite{angrist_2009_mostly}. Since our treatments ($D_i$) are randomized, confounders and evenly spread across control and treatment groups. The error term is then independent to the assignment of treatment ($\epsilon_i \perp\!\!\!\!\perp D_i$), and therefore there is no selection bias. The coefficient $\beta_1$ will then have a causal interpretation and is an unbiased estimate of the causal effect of upweighting on a document's memorization. 

We compute our memorization metrics on up to 1000 documents for each dataset component, in the first shard of the Pile (approximately 3\% of the training set) and test set.
Figure \ref{fig:plot7bmink} contains a graphical representation of $\beta_1$, the estimated causal effect of upweighting on memorization, as measured by MinK\%, for the 7B model. Additional results on upweighting are deferred to Appendix \ref{appendix:upweighting}, including the same plots for other metrics in Figure \ref{fig:all_metrics}, and a regression analysis in Table \ref{tab:mink_regression}.

\paragraph{Results} Based on Figure \ref{fig:plot7bmink}, we conclude that upweighting a document does not substantially affect the memorization of that document for Pythia. Documents exhibited relatively small memorization differences between the control and treatment groups compared to the natural variation across datasets. Figure \ref{fig:all_metrics} shows that this is generally true across memorization metrics. This indicates that slightly duplicating a document had nearly no influence on its memorization. In \citeauthor{nytimes_v_microsoft}, the NYTimes alleged that their data was upsampled during training. However, our causal analysis of Pythia demonstrates that the relationship between upsampling and memorization is not always intuitive and that upsampling does not directly imply memorization. Small amounts of upweighting may not lead to memorization, concurring with \citet{huang-etal-2024-demystifying} which found that the number of times text needs to be seen during training to be memorized is substantial. 

\paragraph{Methodological implications} For Pythia, the train/test split helped us to assign control and treatment groups to make a valid causal comparison. While upweighting a single document is a simple training decision, this method is useful to analyze the effect of training decisions on single pieces of data. For instance, LLama 3 \cite{grattafiori2024llama3herdmodels} uses simulated annealing, where they selected a subset of very high quality data to appear more frequently at the end of training. If Llama3  were on trial, whether a document's inclusion in the annealing set affected its memorization could be analyzed by holding out a train/test split of the annealing data. We recommend that model developers hold out these splits to enable post-hoc reporting on the effects of their training decisions on memorization.

\subsection{A correlational analysis at the dataset level} \label{sec:curation}

\begin{figure}[t]
    \centering
\begin{tikzpicture}[scale=0.85]
    % Define the bounding box for the left half of the page
    \begin{scope}[xshift=-3cm]
        % Draw the circle representing the radius
        \draw[thick, dashed] (0, 0) circle (2);

        % Draw the central data point
        \filldraw[red] (0, 0) circle (2pt) node[anchor=north] {};

        % Draw circular neighbors within the radius
        \filldraw[blue] (1.2, 0.6) circle (2pt) node[anchor=south west] {};
        \filldraw[blue] (-1.2, 0.6) circle (2pt) node[anchor=south east] {};

        % Draw square neighbors within the radius
        \draw[red, thick] (0.6, -1.2) circle (2pt);
        \draw[red, thick] (-0.9, -0.9) circle (2pt);

        % Draw neighbors outside the radius
        \filldraw[gray] (2.5, 0.7) circle (2pt) node[anchor=west] {};
        \filldraw[gray] (-2.8, -0.8) circle (2pt);

        % Add the radius annotation
        \draw[<->, thick] (0, 0) -- (2, 0) node[midway, above] {$r=50$};
    \end{scope}

    % Draw the legend closer to the figure
    \begin{scope}[xshift=3cm]
        \node[anchor=west] at (0, 2) {};
        
        % Central data point
        \filldraw[red] (0, 1.5) circle (2pt);
        \node[anchor=west] at (0.3, 1.5) {StackExchange snippet};
        
        % Circular neighbors within the radius
        \filldraw[blue] (0, 0.5) circle (2pt);
        \node[anchor=west] at (0.3, 0.5) {Snippets from other datasets};
        
        % Square neighbors within the radius
        \draw[red, thick] (0, 1)  circle (2pt);
        \node[anchor=west] at (0.3, 1) {Other StackExchange snippets};
    \end{scope}~\newline~\newline~\newline~\newline~
\end{tikzpicture}
    \caption{We illustrate the intuition of simulated ablation. Under the data density hypothesis, pieces of text are upweighted by other similar text. We fit a regression that predicts the loss of an average i.e. StackExchange snippet given that dataset's typical neighborhood. To simulate ablating a dataset, we predict the loss based on the number of neighbors if that dataset was removed.}
    ~
    \label{fig:intuition}
    \centering
    \includegraphics[]{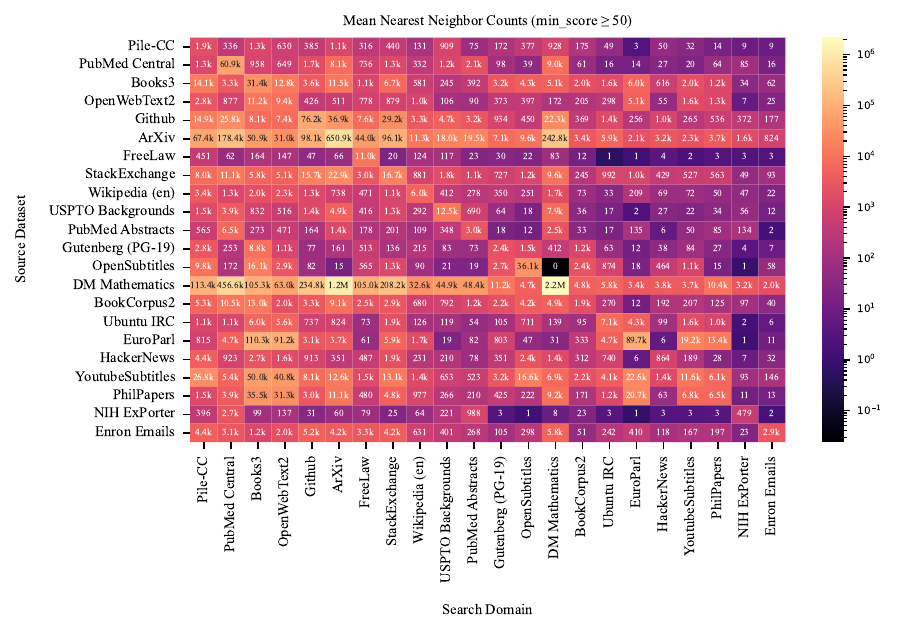}
    \caption{Heatmap of the average neighborhood (BM25 > 50) for snippets from each source dataset (y-axis). Neighbors from each search dataset (x-axis) are retrieved and tallied using Elasticsearch as described in \S\ref{sec:curation}. Darker cells indicate fewer nearest neighbors, and lighter cells indicate more. Note that the table is asymmetrical. Examining the map row-by-row provides the composition of the average neighborhood for each  dataset. Examining the map column-by-column provides the neighborhood contribution for each dataset to the others. With a few exceptions, while most datasets provide the most neighbors to itself, that is often only a small fraction of the total neighbors. }
        \label{fig:dataset_overlaps}
\end{figure}

In the last section, we conducted a causal analysis on the effect of upweighting a single document. However, in legal disputes, particularly class action lawsuits, cases are often litigated on behalf of a class of copyright holders, and the memorization across an entire dataset or a substantial body of work is more relevant. In this section, we study the counterfactual question: how would the memorization of an entire dataset change if it was excluded? We demonstrate how to simulate an ablation study, without having to retrain the LLM. Unlike our previous analysis, this analysis is correlational, but provides insight into which ablation study should be conducted when compute resources are limited. 

\paragraph{Correlational analysis} Unlike our previous analysis on upweighting, we cannot leverage a known source of randomness to create control and treatment groups. Without randomization, we conduct a correlational study, and deriving any causal insight from the analysis requires care. A correlational analysis cannot definitively disentangle whether observed memorization effects are due to implicit weighting or confounding factors such as dataset quality. Despite these limitations, correlational analysis can still provide valuable insights when paired with qualitative reasoning and domain expertise. By identifying patterns of memorization, it serves as a diagnostic tool to identify further areas of  experimentation when computational resources are limited.

\paragraph{Data density} A thread of research has focused on the validity of the data density hypothesis for deep learning, which states that the ability of a model on an example is impacted by the density of the training set around that example \cite{pmlr-v202-kandpal23a, yauney-etal-2023-data, kirchenbauer2024lmd}. The literature here mostly consists of correlational results, as it is difficult to retrain an LLM on data with different data density characteristics. However, the consensus is that data density does play a causal role in memorization \cite{kirchenbauer2024lmd}. Similar to a difference-in-differences analysis, some studies also conduct partial pretraining and find that training on an example or its related texts causes memorization to peak on that example, which decays over training \cite{lesci-etal-2024-causal, huang-etal-2024-demystifying, chang2024largelanguagemodelsacquire}.

\begin{figure}[]
    \centering
    \includegraphics[]{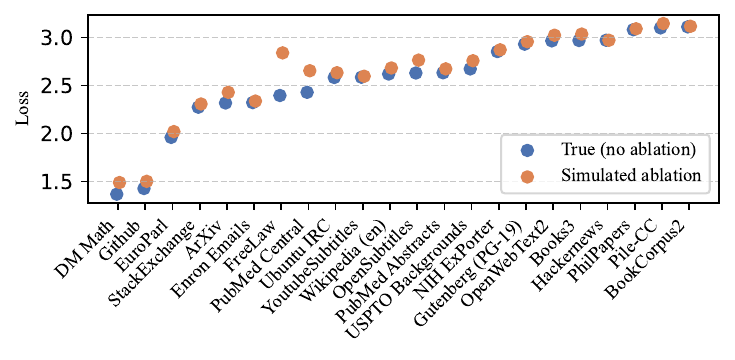}
    \caption{Loss of Pythia 6.9b on each dataset when trained on the entire Pile (blue) or when that dataset is ablated (orange). To simulate an ablation, we first fit a regression to predict the loss of a dataset based on the dataset's average neighborhood. The ablated loss is predicted based on a counterfactual neighborhood count and an illustration is given in Figure \ref{fig:scatter}. While such a regression is not guaranteed to be causal, it can be useful for identifying areas for further experimentation when computational resources are limited. Notably, FreeLaw and PubMed Central are most affected when they are ablated from the training data.}
    \label{fig:ablation_plot}
\end{figure}

\paragraph{Implementing nearest neighbors} \citet{kirchenbauer2024lmd} present a system to estimate training data density by exhaustively embedding snippets of the training text using a neural document embedder and indexing the embeddings in a vector database. The density around a snippet can be calculated by querying for the nearest neighbors of that snippet. Figure \ref{fig:intuition} provides an illustration of a snippet's neighborhood. These derived densities explain Pythia's loss well, but both embedding and indexing are computationally intensive (embedding nearly 1TB of text and indexing several TB of vectors) so we could not reproduce their setup. Instead, we greatly simplify their method by indexing snippets with Elasticsearch, similar to \citet{DBLP:conf/iclr/ElazarBMRSSWGS024}. 

Elasticsearch tokenizes documents by white space, and indexes them according to BM25, a metric similar to tf-idf.\footnote{\url{https://en.wikipedia.org/wiki/Okapi_BM25}} Starting from the beginning of the training data, we index snippets of 50 tokens, taking 40-token strides, resulting in 10 tokens of overlap between consecutive snippets. We query and characterize the neighborhoods of an average snippet from a dataset component. Since we only need to analyze neighbors in aggregate (for each dataset component), indexing only a small sample of the Pile ($4.4\%$) is sufficient, resulting in an index less than 100GB. We experiment with a few different BM25 thresholds in Table \ref{tab:bm25_thresholds}, and find that data density derived from a neighborhood with $r > 50$ approximates model loss well. Table \ref{tab:correlations} shows the correlations between the number of neighbors for the average snippet from each dataset to our memorization metrics. Our Elasticsearch correlates with most memorization metrics well, except for verbatim memorization. Figure \ref{fig:dataset_overlaps} shows a matrix of dataset overlaps, where the rows describe the neighborhood of the average snippet from that dataset. 

\paragraph{Simulating ablations} Since we observe that the number of nearest neighbors correlate with loss well, we regress the neighbors against the average snippet loss of that dataset (depicted in Figure \ref{fig:scatter}). While there is no guarantee such a regression has a causal interpretation, qualitative insights from the data density literature suggests density plays a causal role. We set up the regression as
\begin{equation}
    Y_i = \alpha + \beta_1 \log(N_i) + \epsilon_i
\end{equation}
where $Y_i$ is the average loss of a snippet from dataset $i$, and $N_i = n^i_1 + n^i_2 + \cdots + n^i_{22}$ is the sum of all the nearest neighbors of the average snippet from the $i$-th dataset (sum of the row in Figure \ref{fig:dataset_overlaps}). This regression is therefore fitted on 22 datapoints, one for each dataset component in the Pile. $\alpha$ and $\beta_1$ then relate the number of neighbors to a snippet's loss, and $\epsilon_i$ accounts for the unobserved error. Since the loss $Y_i$ and the number of neighbors $N_i$ are correlated (as opposed to snippets being randomly assigned a desnity $N_i$), confounders may exist, and $\beta_i$ is not guaranteed to be a causal estimate. However, we still use the regression for its predictive value. To estimate the effect of ablating dataset $i$, we can calculate a counterfactual density $N_i'=N_i - n^i_i$ and adjust the loss with
\begin{equation}
    \Delta Y_i = \beta_1 ( \log(N_i) - \log(N_i') )
\end{equation}
which essentially shifts each dataset's loss along a regression line as illustrated in Figure \ref{fig:scatter}. We simulate the loss of each dataset component when ablating that dataset and display the results in Figure \ref{fig:ablation_plot}.

\paragraph{Results} Based on our simulated ablations in Figure \ref{fig:ablation_plot}, we predict that the losses of most datasets would not change much if that dataset was excluded. Looking at our dataset overlaps, Table \ref{fig:dataset_overlaps} shows that, while most datasets, such as Books3 or StackExchange, provide many neighbors to itself, it is often only a small fraction of the total neighbors provided by other datasets. However, some datasets, such as FreeLaw or PubMed Central would have a noticeable change in loss. Here, FreeLaw's neighbors mainly originate from FreeLaw itself, and the removal of those neighbors causes a large change in the predicted, ablated loss.

% can we insert in a dataset so that we can make good comparisons for implicit weighting?
\paragraph{Methodological implications} Interpreting correlational results is challenging, and any derived causal insight should be met with skepticism as it may be based on false comparison. Paired with qualitative reasoning, it is still possible to derive actionable insights. Simulated ablations may still be useful for identifying problematic dataset overlaps warranting deeper investigation. When correlational evidence is the only form of quantitative analysis available, we should pair it with sound causal evidence.  In \citeauthor{kadrey_v_meta}, Meta ran additional ablation studies to justify their choice of training data. The simulated ablations we presented here could help guide such explorations and identify experiments of interest when computational resources are limited.

\section{Formalizing ``fair learning''} \label{sec:formalizing}

Having operationalized ``fair learning'' in \S\ref{sec:deconstruct}, we now formalize our legal reasoning and further theorize a fair learning doctrine. \citeauthor{field2006}, which enshrines the use of robots.txt for web crawling, is an example of how judicial rulemaking rooted in technological decisions extends beyond individual cases and shapes industry norms. Similarly, a fair learning doctrine must be designed with broad societal implications in mind, balancing the interests of innovation and rightholders, while promoting trust and accountability of new technology. By setting an appropriate standard, the doctrine can incentivize design choices that align with ethical and legal norms. Finally, a fair learning doctrine should evolve to provide clarity and practical guidance, establishing a framework for the responsible development of LLMs.

\subsection{A fair learning doctrine}

We formally define fair learning for language models as 
\begin{quote}
\it
    a training regime that does not substantially increase the memorization of copyrighted data.
\end{quote} This definition would frame fair learning around a causal question, which complements the technical methodology we presented in \S\ref{sec:deconstruct} to examine training decisions. Legal doctrines not only structure judicial reasoning but also facilitate the development of social norms \cite{fuller1978forms}. Ideally, a fair learning doctrine achieves these normative goals:

% similar to material contribution and design defects
% forms and limits of adjudication

\paragraph{Encouraging transparency} Courts can play a key role in encouraging transparency for LLMs. To address information asymmetry, courts may shift the burden of proof to model developers or adopt adverse inference (i.e., presuming that undisclosed design choices mean copyright infringement unless evidence is provided to the contrary). By deconstructing design decisions in court and requesting basic memorization analyses, judges signal to model developers the importance of analyzing and documenting the impact of their design choices, beyond simple disclosures such as model cards \cite{mitchell_2019_model}. These mechanisms can incentivize developers to to hold train/test splits for causal analysis (as in \S\ref{sec:upweighting}) and proactively release information about their model’s performance and memorization characteristics, which reduces their litigation risks and fosters transparency.

\paragraph{Promoting responsible innovation} The fair learning doctrine places legal importance on the training decisions of LLMs. By establishing a precedent for fair learning, courts set the expectation that developers should consider the legal and ethical implications of their training practices and actively shape their models to align with these norms. Ideally, developers would be encouraged not only to be compliant but also to innovate on fair learning techniques that reduce memorization and mitigate copyright risks. Fair learning encourages developers to adopt training decisions that are both legally defensible and ethically sound, which fosters responsible innovation.

\paragraph{Balancing competing interests} The fair learning doctrine seeks to balance the interests of copyright holders with the interests of model developers. A key feature of the fair learning doctrine is the threshold used to determine substantial memorization. Courts can adjust this threshold to favor either innovation or rightholder interests. By calibrating this threshold based on the context of each case, courts can flexibly apply fair learning standards to maintain an equilibrium between parties that promote the normative goals of copyright.

% model developers need to ensure that all training decisions are fair, which is hard
% otoh, 

% more coming from the theoretical view
% how will fair learning be evaluated?

% this puts the actions on observation, assuming that there are good and bad actions

% this should only be part of the analysis, and fair learning should be one such consideration in a larger fair use analysis.

% \paragraph{Difficulty of drawing binary rules} After conducting a causal analysis of LLM design decisions, translating these findings into strict binary rules remains challenging. There are many different aspects of LLM memorization and many metrics of interest are continuous \cite{hartmann2023sokmemorizationgeneralpurposelarge, schwarzschild2024rethinkingllmmemorizationlens}. As discussed in \S\ref{sec:curation}, including  StackExchange in the training data significantly influenced Github memorization, yet defining a precise threshold for excessive memorization is difficult. Judging from the results, there were no clear signs of excessive implicit upweighting. Standards provide courts with the flexibility to make nuanced assessments, sidestepping the need to make numerical cutoffs in legal decisions.

% what are the normative goals that fair learning should try to enable?
% perhaps fair learning allows companies to be more transparent

\subsection{Litigating fair learning}

% practical epistemic considerations, on how difficult it is to prove what is fair learning.
We now elaborate on the application of fair learning in litigating a copyright infringement case: 

\paragraph{1. Establishing prima facie infringement} The plaintiff first establishes a prima facie case, demonstrating ownership of the copyrighted work and unauthorized copying during the training of the defendant's LLM. Since details of LLM training data are rarely disclosed \cite{liesenfeld_rethinking_2024},  courts can (and do \cite{masterlist_ai_lawsuits_2024}) infer copying from instances of substantial similarity or verbatim reproduction in the model outputs, along with evidence of access to the copyrighted material through documentation or membership inference techniques \cite{wei-etal-2024-proving}. 

\paragraph{2. Invoking the fair learning doctrine} Once a prima facie case of infringement is established, the court may invoke the fair learning doctrine to assess whether the defendant’s training decisions substantially contributed to the model’s memorization of copyrighted material. However, analyzing training decisions requires access to the model, training details and data, none of which the court or plaintiff have access to. There is also an asymmetry in expertise, as those capable of conducting the technical analyses are likely to be the model developers rather than judges or plaintiffs. In this context, information asymmetry impedes the plaintiff's ability to establish their claims which hinders a fair litigation process. Courts can play an important role in promoting procedural fairness by shifting the burden of proof to the model developers \cite{kaplow2011burden} or adopting adverse inference, and they should also seek to leverage the defendant's private expertise to rigorously analyze training decisions \cite{selbst2021institutional}.

\paragraph{3. Defendant's rebuttal} Now that the defendant bears the burden of proof, they must demonstrate that their training decisions did not substantially contribute to memorization. Ideally, defendants would already have documentation, ablation studies, and memorization analyses prepared as part of their internal development. Once the defendant releases training details and provides an initial analysis, the burden of proof shifts back to the plaintiff to identify unfair training decisions or gaps in the defendant’s analysis. This shift ensures fairness, as it is impractical for defendants to prove that every training decision complies with fair learning, whereas plaintiffs are better positioned to highlight specific concerns or omissions. At the same time, courts need to consider the quality of the statistical evidence presented. In the best case, a causal analysis such as the one in \S\ref{sec:upweighting} is sufficient. However, some decisions may be harder to analyze and only allow for correlational analysis, such as in \S\ref{sec:curation}. In this case, courts may request additional experimentation to clarify the evidence, which is consistent with the recent approach in \citeauthor{kadrey_v_meta}, where Meta conducted additional ablation experiments to address evidentiary gaps.

\paragraph{4. Applying the substantiality threshold} Once the relevant measurements are made, courts must determine how much memorization is too much and whether the model in question substantially memorizes its training data. Here, courts should consider the broader normative goals of copyright law and set an appropriate threshold which balances the interests of model developers against copyright holders. The evaluation of substantiality in our proposed fair learning doctrine will be a facts-and-circumstances standard tailored to each case, leaving room for slippage and potential inconsistency. It may therefore be desirable to develop more specific guidelines for what exactly constitutes substantiality.

\subsection{Enhancing the doctrine's clarity}

Our formulation of fair learning emphasizes the use of legal standards rather than rules. Standards require interpretation after the fact (e.g., deciding whether a driver followed a law requiring them to "drive at a reasonable speed"), whereas rules specify how actions will be judged in advance (e.g., "the speed limit is 55 mph"). Since the risks and impacts of emerging technology are uncertain, courts frequently rely on standards, which allow for flexibility and adaptability when a dispute arises. The merits of standards versus rules have long been debated and \citet{kaplow1992rules} observes their economic difference: since rules are specified before the fact, they require more effort to set but reduce the burden of interpretation. On the other hand, standards require less effort to set but shift the burden to those interpreting it.

Rules would provide much needed clarity for both model developers and copyright holders by delineating legal boundaries. By specifying when LLM training constitutes infringement, rules would foster innovation within these boundaries and encourage the ethical use of copyrighted material. Creating safe harbors, such as those included in the Digital Millennium Copyright Act (DMCA), will require us to theorize about good rules \cite{lemley2017rationalizing}. Although the DMCA safe harbors are not without flaws, a good safe harbor would establish clear responsibilities and protections for model developers, enabling them to operate without the constant threat of litigation. However setting rules can pose significant challenges \cite{schauer_thinking_2009}. Once a rule is set, it risks being over- or under-inclusive \cite{burk2005legal}. Rules can also quickly become obsolete and fail to account for evolving technological structures \cite{yew_break_2024}. 

Clarifying the fair learning doctrine would require making it more rule-like. Because judges lack technical expertise, one common---and perhaps necessary---approach is to defer to external technical guidelines (e.g. fair use in \citeauthor{field2006} partly depended on adhering to industry best practices). In areas such as cybersecurity \cite{shackelford2015global} and environmental regulation \cite{wagner2000triumph} courts can defer to external technical guidelines, including those from regulatory agencies (e.g. EPA), standards-setting bodies (e.g. NIST), or private organizations (e.g. IEEE). By relying on external guidelines, courts leverage specialized expertise and foster a holistic approach to governance \cite{marchant2020governance}. Encouraging the development of a consistent body of ``soft'' law could temper inconsistency in the substantiality threshold of fair learning \cite{hagemann2018soft}. However, reliance on external guidelines raises concerns about accountability and democratic legitimacy. Private organizations and regulatory agencies may not always be transparent or publicly accountable, mirroring longstanding debates in administrative law \cite{araiza2022short}. Over-reliance on industry-driven guidelines also risks regulatory capture, where dominant firms co-opt standards to serve their interests rather than the public good \cite{young_confronting_2022}.

\section{Conclusion}
In this work, we explore a structural perspective on technology to address copyright issues for LLMs. This presents an opportunity to formalize a notion of ``fair learning'' and refine copyright law for future challenges \cite{lemley2020fair}. We operationalize "fair learning" in terms of substantial memorization and demonstrate statistical methods to assess the impact of design decisions on model memorization. By theorizing a fair learning doctrine, we highlight that courts face challenges in information asymmetry, quality of statistical evidence, and determining a substantiality threshold. We provide a number of actionable insights to better enable courts to advance copyright's normative goals, and hope to offer a useful starting point in formulating a fair learning doctrine and clarifying acceptable LLM design. 

Courts should not shy away from deconstructing technology and refining legal doctrine. The impact of training decisions on LLM memorization can be measured and optimized, and machine learning research can play an important role by providing empirical guidance on copyright-safe design. By deferring to external expertise, there is opportunity to open up a holistic governance approach, but this also carries risks, including regulatory capture and lack of public accountability. Copyright is only the first legal frontier for LLMs. As we grapple with the broader societal impacts of these models, we hope that the legal frameworks established here will illuminate the legal options for other areas.
%%
%% The acknowledgments section is defined using the "acks" environment
%% (and NOT an unnumbered section). This ensures the proper
%% identification of the section in the article metadata, and the
%% consistent spelling of the heading.
% \begin{acks}
% Acknowledgments.
% \end{acks}

%%
%% The next two lines define the bibliography style to be used, and
%% the bibliography file.
\bibliographystyle{ACM-Reference-Format}
\bibliography{sample-base}

%%% -*-BibTeX-*-
%%% Do NOT edit. File created by BibTeX with style
%%% ACM-Reference-Format-Journals [18-Jan-2012].

\begin{thebibliography}{90}

%%% ====================================================================
%%% NOTE TO THE USER: you can override these defaults by providing
%%% customized versions of any of these macros before the \bibliography
%%% command.  Each of them MUST provide its own final punctuation,
%%% except for \shownote{}, \showDOI{}, and \showURL{}.  The latter two
%%% do not use final punctuation, in order to avoid confusing it with
%%% the Web address.
%%%
%%% To suppress output of a particular field, define its macro to expand
%%% to an empty string, or better, \unskip, like this:
%%%
%%% \newcommand{\showDOI}[1]{\unskip}   % LaTeX syntax
%%%
%%% \def \showDOI #1{\unskip}           % plain TeX syntax
%%%
%%% ====================================================================

\ifx \showCODEN    \undefined \def \showCODEN     #1{\unskip}     \fi
\ifx \showDOI      \undefined \def \showDOI       #1{#1}\fi
\ifx \showISBNx    \undefined \def \showISBNx     #1{\unskip}     \fi
\ifx \showISBNxiii \undefined \def \showISBNxiii  #1{\unskip}     \fi
\ifx \showISSN     \undefined \def \showISSN      #1{\unskip}     \fi
\ifx \showLCCN     \undefined \def \showLCCN      #1{\unskip}     \fi
\ifx \shownote     \undefined \def \shownote      #1{#1}          \fi
\ifx \showarticletitle \undefined \def \showarticletitle #1{#1}   \fi
\ifx \showURL      \undefined \def \showURL       {\relax}        \fi
% The following commands are used for tagged output and should be
% invisible to TeX
\providecommand\bibfield[2]{#2}
\providecommand\bibinfo[2]{#2}
\providecommand\natexlab[1]{#1}
\providecommand\showeprint[2][]{arXiv:#2}

\bibitem[Angrist and Pischke(2009)]%
        {angrist_2009_mostly}
\bibfield{author}{\bibinfo{person}{Joshua~D. Angrist} {and} \bibinfo{person}{Jörn-Steffen Pischke}.} \bibinfo{year}{2009}\natexlab{}.
\newblock \bibinfo{booktitle}{\emph{Mostly Harmless Econometrics: An Empiricist's Companion}}.
\newblock \bibinfo{publisher}{Princeton University Press}.
\newblock
\showISBNx{9780691120348}
\urldef\tempurl%
\url{http://www.jstor.org/stable/j.ctvcm4j72}
\showURL{%
\tempurl}


\bibitem[Araiza(2022)]%
        {araiza2022short}
\bibfield{author}{\bibinfo{person}{William~D. Araiza}.} \bibinfo{year}{2022}\natexlab{}.
\newblock \bibinfo{booktitle}{\emph{A Short \& Happy Guide to Administrative Law} (\bibinfo{edition}{2nd} ed.)}.
\newblock \bibinfo{publisher}{West Academic Publishing}.
\newblock


\bibitem[Atkinson and Morrison(2024)]%
        {atkinson2024legalrisktaxonomygenerative}
\bibfield{author}{\bibinfo{person}{David Atkinson} {and} \bibinfo{person}{Jacob Morrison}.} \bibinfo{year}{2024}\natexlab{}.
\newblock \bibinfo{title}{A Legal Risk Taxonomy for Generative Artificial Intelligence}.
\newblock
\newblock
\showeprint[arxiv]{2404.09479}~[cs.CY]
\urldef\tempurl%
\url{https://arxiv.org/abs/2404.09479}
\showURL{%
\tempurl}


\bibitem[Biderman et~al\mbox{.}(2023)]%
        {DBLP:conf/icml/BidermanSABOHKP23}
\bibfield{author}{\bibinfo{person}{Stella Biderman}, \bibinfo{person}{Hailey Schoelkopf}, \bibinfo{person}{Quentin~Gregory Anthony}, \bibinfo{person}{Herbie Bradley}, \bibinfo{person}{Kyle O'Brien}, \bibinfo{person}{Eric Hallahan}, \bibinfo{person}{Mohammad~Aflah Khan}, \bibinfo{person}{Shivanshu Purohit}, \bibinfo{person}{USVSN~Sai Prashanth}, \bibinfo{person}{Edward Raff}, \bibinfo{person}{Aviya Skowron}, \bibinfo{person}{Lintang Sutawika}, {and} \bibinfo{person}{Oskar van~der Wal}.} \bibinfo{year}{2023}\natexlab{}.
\newblock \showarticletitle{Pythia: {A} Suite for Analyzing Large Language Models Across Training and Scaling}. In \bibinfo{booktitle}{\emph{International Conference on Machine Learning, {ICML} 2023, 23-29 July 2023, Honolulu, Hawaii, {USA}}} \emph{(\bibinfo{series}{Proceedings of Machine Learning Research}, Vol.~\bibinfo{volume}{202})}, \bibfield{editor}{\bibinfo{person}{Andreas Krause}, \bibinfo{person}{Emma Brunskill}, \bibinfo{person}{Kyunghyun Cho}, \bibinfo{person}{Barbara Engelhardt}, \bibinfo{person}{Sivan Sabato}, {and} \bibinfo{person}{Jonathan Scarlett}} (Eds.). \bibinfo{publisher}{{PMLR}}, \bibinfo{pages}{2397--2430}.
\newblock
\urldef\tempurl%
\url{https://proceedings.mlr.press/v202/biderman23a.html}
\showURL{%
\tempurl}


\bibitem[Birhane et~al\mbox{.}(2022)]%
        {birhane_2022_values}
\bibfield{author}{\bibinfo{person}{Abeba Birhane}, \bibinfo{person}{Pratyusha Kalluri}, \bibinfo{person}{Dallas Card}, \bibinfo{person}{William Agnew}, \bibinfo{person}{Ravit Dotan}, {and} \bibinfo{person}{Michelle Bao}.} \bibinfo{year}{2022}\natexlab{}.
\newblock \showarticletitle{The Values Encoded in Machine Learning Research}. In \bibinfo{booktitle}{\emph{Proceedings of the 2022 ACM Conference on Fairness, Accountability, and Transparency}} (Seoul, Republic of Korea) \emph{(\bibinfo{series}{FAccT '22})}. \bibinfo{publisher}{Association for Computing Machinery}, \bibinfo{address}{New York, NY, USA}, \bibinfo{pages}{173–184}.
\newblock
\showISBNx{9781450393522}
\urldef\tempurl%
\url{https://doi.org/10.1145/3531146.3533083}
\showDOI{\tempurl}


\bibitem[Bommasani et~al\mbox{.}(2021)]%
        {Bommasani2021FoundationModels}
\bibfield{author}{\bibinfo{person}{Rishi Bommasani} {et~al\mbox{.}}} \bibinfo{year}{2021}\natexlab{}.
\newblock \showarticletitle{On the Opportunities and Risks of Foundation Models}.
\newblock \bibinfo{journal}{\emph{ArXiv}} (\bibinfo{year}{2021}).
\newblock
\urldef\tempurl%
\url{https://crfm.stanford.edu/assets/report.pdf}
\showURL{%
\tempurl}


\bibitem[Brittain(2024)]%
        {brittain2024}
\bibfield{author}{\bibinfo{person}{Blake Brittain}.} \bibinfo{year}{2024}\natexlab{}.
\newblock \showarticletitle{OpenAI says New York Times hacked ChatGPT to build copyright lawsuit}.
\newblock \bibinfo{journal}{\emph{Reuters Technology}} (\bibinfo{date}{27 February} \bibinfo{year}{2024}).
\newblock
\urldef\tempurl%
\url{https://www.reuters.com/technology/cybersecurity/openai-says-new-york-times-hacked-chatgpt-build-copyright-lawsuit-2024-02-27/}
\showURL{%
\tempurl}
\newblock
\shownote{Accessed: 2025-01-22}.


\bibitem[Burk(2005)]%
        {burk2005legal}
\bibfield{author}{\bibinfo{person}{Dan~L. Burk}.} \bibinfo{year}{2005}\natexlab{}.
\newblock \showarticletitle{Legal and Technical Standards in Digital Rights Management Technology}.
\newblock \bibinfo{journal}{\emph{Fordham Law Review}}  \bibinfo{volume}{74} (\bibinfo{year}{2005}), \bibinfo{pages}{537}.
\newblock
\urldef\tempurl%
\url{https://ir.lawnet.fordham.edu/flr/vol74/iss2/9}
\showURL{%
\tempurl}


\bibitem[Carlini et~al\mbox{.}(2019)]%
        {carlini2019}
\bibfield{author}{\bibinfo{person}{Nicholas Carlini}, \bibinfo{person}{Chang Liu}, \bibinfo{person}{Jeremiah Kos}, \bibinfo{person}{{\'U}lfar Erlingsson}, {and} \bibinfo{person}{Dawn Song}.} \bibinfo{year}{2019}\natexlab{}.
\newblock \showarticletitle{The Secret Sharer: Evaluating and Testing Unintended Memorization in Neural Networks}. In \bibinfo{booktitle}{\emph{28th {USENIX} Security Symposium ({USENIX} Security 19)}}. USENIX Association, \bibinfo{pages}{267--284}.
\newblock


\bibitem[Carlini et~al\mbox{.}(2021)]%
        {carlini_extracting_2021}
\bibfield{author}{\bibinfo{person}{Nicholas Carlini}, \bibinfo{person}{Florian Tram{\`e}r}, \bibinfo{person}{Eric Wallace}, \bibinfo{person}{Matthew Jagielski}, \bibinfo{person}{Ariel Herbert-Voss}, \bibinfo{person}{Katherine Lee}, \bibinfo{person}{Adam Roberts}, \bibinfo{person}{Tom Brown}, \bibinfo{person}{Dawn Song}, \bibinfo{person}{{\'U}lfar Erlingsson}, \bibinfo{person}{Alina Oprea}, {and} \bibinfo{person}{Colin Raffel}.} \bibinfo{year}{2021}\natexlab{}.
\newblock \showarticletitle{Extracting Training Data from Large Language Models}. In \bibinfo{booktitle}{\emph{30th USENIX Security Symposium (USENIX Security 21)}}. \bibinfo{publisher}{USENIX Association}, \bibinfo{pages}{2633--2650}.
\newblock
\showISBNx{978-1-939133-24-3}
\urldef\tempurl%
\url{https://www.usenix.org/conference/usenixsecurity21/presentation/carlini-extracting}
\showURL{%
\tempurl}


\bibitem[Carlini et~al\mbox{.}(2020)]%
        {Carlini2020ExtractingTD}
\bibfield{author}{\bibinfo{person}{Nicholas Carlini}, \bibinfo{person}{Florian Tram{\`e}r}, \bibinfo{person}{Eric Wallace}, \bibinfo{person}{Matthew Jagielski}, \bibinfo{person}{Ariel Herbert-Voss}, \bibinfo{person}{Katherine Lee}, \bibinfo{person}{Adam Roberts}, \bibinfo{person}{Tom~B. Brown}, \bibinfo{person}{Dawn~Xiaodong Song}, \bibinfo{person}{{\'U}lfar Erlingsson}, \bibinfo{person}{Alina Oprea}, {and} \bibinfo{person}{Colin Raffel}.} \bibinfo{year}{2020}\natexlab{}.
\newblock \showarticletitle{Extracting Training Data from Large Language Models}. In \bibinfo{booktitle}{\emph{USENIX Security Symposium}}.
\newblock
\urldef\tempurl%
\url{https://api.semanticscholar.org/CorpusID:229156229}
\showURL{%
\tempurl}


\bibitem[Chang et~al\mbox{.}(2024)]%
        {chang2024largelanguagemodelsacquire}
\bibfield{author}{\bibinfo{person}{Hoyeon Chang}, \bibinfo{person}{Jinho Park}, \bibinfo{person}{Seonghyeon Ye}, \bibinfo{person}{Sohee Yang}, \bibinfo{person}{Youngkyung Seo}, \bibinfo{person}{Du-Seong Chang}, {and} \bibinfo{person}{Minjoon Seo}.} \bibinfo{year}{2024}\natexlab{}.
\newblock \bibinfo{title}{How Do Large Language Models Acquire Factual Knowledge During Pretraining?}
\newblock
\newblock
\showeprint[arxiv]{2406.11813}~[cs.CL]
\urldef\tempurl%
\url{https://arxiv.org/abs/2406.11813}
\showURL{%
\tempurl}


\bibitem[Chang et~al\mbox{.}(2023)]%
        {chang-etal-2023-speak}
\bibfield{author}{\bibinfo{person}{Kent Chang}, \bibinfo{person}{Mackenzie Cramer}, \bibinfo{person}{Sandeep Soni}, {and} \bibinfo{person}{David Bamman}.} \bibinfo{year}{2023}\natexlab{}.
\newblock \showarticletitle{Speak, Memory: An Archaeology of Books Known to {C}hat{GPT}/{GPT}-4}. In \bibinfo{booktitle}{\emph{Proceedings of the 2023 Conference on Empirical Methods in Natural Language Processing}}, \bibfield{editor}{\bibinfo{person}{Houda Bouamor}, \bibinfo{person}{Juan Pino}, {and} \bibinfo{person}{Kalika Bali}} (Eds.). \bibinfo{publisher}{Association for Computational Linguistics}, \bibinfo{address}{Singapore}, \bibinfo{pages}{7312--7327}.
\newblock
\urldef\tempurl%
\url{https://doi.org/10.18653/v1/2023.emnlp-main.453}
\showDOI{\tempurl}


\bibitem[{ChatGPT is Eating the World}(2024)]%
        {masterlist_ai_lawsuits_2024}
\bibfield{author}{\bibinfo{person}{{ChatGPT is Eating the World}}.} \bibinfo{year}{2024}\natexlab{}.
\newblock \bibinfo{title}{Master List of Lawsuits v. AI: ChatGPT, OpenAI, Microsoft, Meta, MidJourney, Other AI Cos.}
\newblock
\newblock
\urldef\tempurl%
\url{https://chatgptiseatingtheworld.com/2024/08/27/master-list-of-lawsuits-v-ai-chatgpt-openai-microsoft-meta-midjourney-other-ai-cos/}
\showURL{%
\tempurl}
\newblock
\shownote{Accessed: 2024-12-25}.


\bibitem[Chen et~al\mbox{.}(2024)]%
        {chen-etal-2024-copybench}
\bibfield{author}{\bibinfo{person}{Tong Chen}, \bibinfo{person}{Akari Asai}, \bibinfo{person}{Niloofar Mireshghallah}, \bibinfo{person}{Sewon Min}, \bibinfo{person}{James Grimmelmann}, \bibinfo{person}{Yejin Choi}, \bibinfo{person}{Hannaneh Hajishirzi}, \bibinfo{person}{Luke Zettlemoyer}, {and} \bibinfo{person}{Pang~Wei Koh}.} \bibinfo{year}{2024}\natexlab{}.
\newblock \showarticletitle{{C}opy{B}ench: Measuring Literal and Non-Literal Reproduction of Copyright-Protected Text in Language Model Generation}. In \bibinfo{booktitle}{\emph{Proceedings of the 2024 Conference on Empirical Methods in Natural Language Processing}}, \bibfield{editor}{\bibinfo{person}{Yaser Al-Onaizan}, \bibinfo{person}{Mohit Bansal}, {and} \bibinfo{person}{Yun-Nung Chen}} (Eds.). \bibinfo{publisher}{Association for Computational Linguistics}, \bibinfo{address}{Miami, Florida, USA}, \bibinfo{pages}{15134--15158}.
\newblock
\urldef\tempurl%
\url{https://doi.org/10.18653/v1/2024.emnlp-main.844}
\showDOI{\tempurl}


\bibitem[Cooper et~al\mbox{.}(2024)]%
        {cooper2024machineunlearningdoesntthink}
\bibfield{author}{\bibinfo{person}{A.~Feder Cooper}, \bibinfo{person}{Christopher~A. Choquette-Choo}, \bibinfo{person}{Miranda Bogen}, \bibinfo{person}{Matthew Jagielski}, \bibinfo{person}{Katja Filippova}, \bibinfo{person}{Ken~Ziyu Liu}, \bibinfo{person}{Alexandra Chouldechova}, \bibinfo{person}{Jamie Hayes}, \bibinfo{person}{Yangsibo Huang}, \bibinfo{person}{Niloofar Mireshghallah}, \bibinfo{person}{Ilia Shumailov}, \bibinfo{person}{Eleni Triantafillou}, \bibinfo{person}{Peter Kairouz}, \bibinfo{person}{Nicole Mitchell}, \bibinfo{person}{Percy Liang}, \bibinfo{person}{Daniel~E. Ho}, \bibinfo{person}{Yejin Choi}, \bibinfo{person}{Sanmi Koyejo}, \bibinfo{person}{Fernando Delgado}, \bibinfo{person}{James Grimmelmann}, \bibinfo{person}{Vitaly Shmatikov}, \bibinfo{person}{Christopher~De Sa}, \bibinfo{person}{Solon Barocas}, \bibinfo{person}{Amy Cyphert}, \bibinfo{person}{Mark Lemley}, \bibinfo{person}{danah boyd}, \bibinfo{person}{Jennifer~Wortman Vaughan}, \bibinfo{person}{Miles Brundage},
  \bibinfo{person}{David Bau}, \bibinfo{person}{Seth Neel}, \bibinfo{person}{Abigail~Z. Jacobs}, \bibinfo{person}{Andreas Terzis}, \bibinfo{person}{Hanna Wallach}, \bibinfo{person}{Nicolas Papernot}, {and} \bibinfo{person}{Katherine Lee}.} \bibinfo{year}{2024}\natexlab{}.
\newblock \bibinfo{title}{Machine Unlearning Doesn't Do What You Think: Lessons for Generative AI Policy, Research, and Practice}.
\newblock
\newblock
\showeprint[arxiv]{2412.06966}~[cs.LG]
\urldef\tempurl%
\url{https://arxiv.org/abs/2412.06966}
\showURL{%
\tempurl}


\bibitem[Crootof and Ard(2021)]%
        {CrootofArd2021}
\bibfield{author}{\bibinfo{person}{Rebecca Crootof} {and} \bibinfo{person}{B.J. Ard}.} \bibinfo{year}{2021}\natexlab{}.
\newblock \showarticletitle{Structuring Techlaw}.
\newblock \bibinfo{journal}{\emph{Harvard Journal of Law \& Technology}}  \bibinfo{volume}{34} (\bibinfo{year}{2021}), \bibinfo{pages}{347}.
\newblock
\urldef\tempurl%
\url{https://doi.org/10.2139/ssrn.3664124}
\showDOI{\tempurl}
\newblock
\shownote{Univ. of Wisconsin Legal Studies Research Paper No. 1761}.


\bibitem[Ding(2023)]%
        {ding2023coursecausalinference}
\bibfield{author}{\bibinfo{person}{Peng Ding}.} \bibinfo{year}{2023}\natexlab{}.
\newblock \bibinfo{title}{A First Course in Causal Inference}.
\newblock
\newblock
\showeprint[arxiv]{2305.18793}~[stat.ME]
\urldef\tempurl%
\url{https://arxiv.org/abs/2305.18793}
\showURL{%
\tempurl}


\bibitem[Elazar et~al\mbox{.}(2024)]%
        {DBLP:conf/iclr/ElazarBMRSSWGS024}
\bibfield{author}{\bibinfo{person}{Yanai Elazar}, \bibinfo{person}{Akshita Bhagia}, \bibinfo{person}{Ian Magnusson}, \bibinfo{person}{Abhilasha Ravichander}, \bibinfo{person}{Dustin Schwenk}, \bibinfo{person}{Alane Suhr}, \bibinfo{person}{Evan~Pete Walsh}, \bibinfo{person}{Dirk Groeneveld}, \bibinfo{person}{Luca Soldaini}, \bibinfo{person}{Sameer Singh}, \bibinfo{person}{Hannaneh Hajishirzi}, \bibinfo{person}{Noah~A. Smith}, {and} \bibinfo{person}{Jesse Dodge}.} \bibinfo{year}{2024}\natexlab{}.
\newblock \showarticletitle{What's In My Big Data?}. In \bibinfo{booktitle}{\emph{The Twelfth International Conference on Learning Representations, {ICLR} 2024, Vienna, Austria, May 7-11, 2024}}. \bibinfo{publisher}{OpenReview.net}.
\newblock
\urldef\tempurl%
\url{https://openreview.net/forum?id=RvfPnOkPV4}
\showURL{%
\tempurl}


\bibitem[Elkin-Koren(2017)]%
        {elkin-koren2017fairuse}
\bibfield{author}{\bibinfo{person}{Niva Elkin-Koren}.} \bibinfo{year}{2017}\natexlab{}.
\newblock \showarticletitle{Fair Use by Design}.
\newblock \bibinfo{journal}{\emph{UCLA Law Review}}  \bibinfo{volume}{64} (\bibinfo{year}{2017}), \bibinfo{pages}{22}.
\newblock
\urldef\tempurl%
\url{https://ssrn.com/abstract=3217839}
\showURL{%
\tempurl}


\bibitem[Elkin-Koren et~al\mbox{.}(2024)]%
        {elkinkoren2024copyrightreducedprivacy}
\bibfield{author}{\bibinfo{person}{Niva Elkin-Koren}, \bibinfo{person}{Uri Hacohen}, \bibinfo{person}{Roi Livni}, {and} \bibinfo{person}{Shay Moran}.} \bibinfo{year}{2024}\natexlab{}.
\newblock \bibinfo{title}{Can Copyright be Reduced to Privacy?}
\newblock
\newblock
\showeprint[arxiv]{2305.14822}~[cs.LG]
\urldef\tempurl%
\url{https://arxiv.org/abs/2305.14822}
\showURL{%
\tempurl}


\bibitem[Feldman(2020)]%
        {vitaly_2020_does}
\bibfield{author}{\bibinfo{person}{Vitaly Feldman}.} \bibinfo{year}{2020}\natexlab{}.
\newblock \showarticletitle{Does learning require memorization? a short tale about a long tail}. In \bibinfo{booktitle}{\emph{Proceedings of the 52nd Annual ACM SIGACT Symposium on Theory of Computing}} (Chicago, IL, USA) \emph{(\bibinfo{series}{STOC 2020})}. \bibinfo{publisher}{Association for Computing Machinery}, \bibinfo{address}{New York, NY, USA}, \bibinfo{pages}{954–959}.
\newblock
\showISBNx{9781450369794}
\urldef\tempurl%
\url{https://doi.org/10.1145/3357713.3384290}
\showDOI{\tempurl}


\bibitem[Franceschelli and Musolesi(2022)]%
        {Franceschelli_Musolesi_2022}
\bibfield{author}{\bibinfo{person}{Giorgio Franceschelli} {and} \bibinfo{person}{Mirco Musolesi}.} \bibinfo{year}{2022}\natexlab{}.
\newblock \showarticletitle{Copyright in generative deep learning}.
\newblock \bibinfo{journal}{\emph{Data \& Policy}}  \bibinfo{volume}{4} (\bibinfo{year}{2022}), \bibinfo{pages}{e17}.
\newblock
\urldef\tempurl%
\url{https://doi.org/10.1017/dap.2022.10}
\showDOI{\tempurl}


\bibitem[Franceschelli and Musolesi(2024)]%
        {Franceschelli2024}
\bibfield{author}{\bibinfo{person}{Giorgio Franceschelli} {and} \bibinfo{person}{Mirco Musolesi}.} \bibinfo{year}{2024}\natexlab{}.
\newblock \showarticletitle{On the creativity of large language models}.
\newblock \bibinfo{journal}{\emph{AI {\&} SOCIETY}} (\bibinfo{date}{28 Nov} \bibinfo{year}{2024}).
\newblock
\showISSN{1435-5655}
\urldef\tempurl%
\url{https://doi.org/10.1007/s00146-024-02127-3}
\showDOI{\tempurl}


\bibitem[Freeman et~al\mbox{.}(2024)]%
        {freeman2024exploring}
\bibfield{author}{\bibinfo{person}{Joshua Freeman}, \bibinfo{person}{Chloe Rippe}, \bibinfo{person}{Edoardo Debenedetti}, {and} \bibinfo{person}{Maksym Andriushchenko}.} \bibinfo{year}{2024}\natexlab{}.
\newblock \showarticletitle{Exploring Memorization and Copyright Violation in Frontier {LLM}s: A Study of the New York Times v. Open{AI} 2023 Lawsuit}. In \bibinfo{booktitle}{\emph{Neurips Safe Generative AI Workshop 2024}}.
\newblock
\urldef\tempurl%
\url{https://openreview.net/forum?id=C66DBl9At8}
\showURL{%
\tempurl}


\bibitem[Fuller(1978)]%
        {fuller1978forms}
\bibfield{author}{\bibinfo{person}{Lon~L. Fuller}.} \bibinfo{year}{1978}\natexlab{}.
\newblock \showarticletitle{{The Forms and Limits of Adjudication}}.
\newblock \bibinfo{journal}{\emph{Harvard Law Review}} \bibinfo{volume}{92}, \bibinfo{number}{2} (\bibinfo{year}{1978}), \bibinfo{pages}{353--409}.
\newblock
\urldef\tempurl%
\url{https://doi.org/10.2307/1340368}
\showDOI{\tempurl}


\bibitem[Gao et~al\mbox{.}(2020)]%
        {pile}
\bibfield{author}{\bibinfo{person}{Leo Gao}, \bibinfo{person}{Stella Biderman}, \bibinfo{person}{Sid Black}, \bibinfo{person}{Laurence Golding}, \bibinfo{person}{Travis Hoppe}, \bibinfo{person}{Charles Foster}, \bibinfo{person}{Jason Phang}, \bibinfo{person}{Horace He}, \bibinfo{person}{Anish Thite}, \bibinfo{person}{Noa Nabeshima}, \bibinfo{person}{Shawn Presser}, {and} \bibinfo{person}{Connor Leahy}.} \bibinfo{year}{2020}\natexlab{}.
\newblock \showarticletitle{The {P}ile: An 800GB Dataset of Diverse Text for Language Modeling}.
\newblock \bibinfo{journal}{\emph{arXiv preprint arXiv:2101.00027}} (\bibinfo{year}{2020}).
\newblock


\bibitem[Grattafiori et~al\mbox{.}(2024)]%
        {grattafiori2024llama3herdmodels}
\bibfield{author}{\bibinfo{person}{Aaron Grattafiori} {et~al\mbox{.}}} \bibinfo{year}{2024}\natexlab{}.
\newblock \bibinfo{title}{The Llama 3 Herd of Models}.
\newblock
\newblock
\showeprint[arxiv]{2407.21783}~[cs.AI]
\urldef\tempurl%
\url{https://arxiv.org/abs/2407.21783}
\showURL{%
\tempurl}


\bibitem[Greenberg(2016)]%
        {greenberg2016technology}
\bibfield{author}{\bibinfo{person}{Brad~A. Greenberg}.} \bibinfo{year}{2016}\natexlab{}.
\newblock \showarticletitle{Rethinking Technology Neutrality}.
\newblock \bibinfo{journal}{\emph{Minnesota Law Review}}  \bibinfo{volume}{100} (\bibinfo{date}{March 16} \bibinfo{year}{2016}), \bibinfo{pages}{1495}.
\newblock
\newblock
\shownote{Available at SSRN: \url{https://ssrn.com/abstract=2748932}}.


\bibitem[Gururangan et~al\mbox{.}(2022)]%
        {gururangan-etal-2022-whose}
\bibfield{author}{\bibinfo{person}{Suchin Gururangan}, \bibinfo{person}{Dallas Card}, \bibinfo{person}{Sarah Dreier}, \bibinfo{person}{Emily Gade}, \bibinfo{person}{Leroy Wang}, \bibinfo{person}{Zeyu Wang}, \bibinfo{person}{Luke Zettlemoyer}, {and} \bibinfo{person}{Noah~A. Smith}.} \bibinfo{year}{2022}\natexlab{}.
\newblock \showarticletitle{Whose Language Counts as High Quality? Measuring Language Ideologies in Text Data Selection}. In \bibinfo{booktitle}{\emph{Proceedings of the 2022 Conference on Empirical Methods in Natural Language Processing}}, \bibfield{editor}{\bibinfo{person}{Yoav Goldberg}, \bibinfo{person}{Zornitsa Kozareva}, {and} \bibinfo{person}{Yue Zhang}} (Eds.). \bibinfo{publisher}{Association for Computational Linguistics}, \bibinfo{address}{Abu Dhabi, United Arab Emirates}, \bibinfo{pages}{2562--2580}.
\newblock
\urldef\tempurl%
\url{https://doi.org/10.18653/v1/2022.emnlp-main.165}
\showDOI{\tempurl}


\bibitem[Hacohen et~al\mbox{.}(2024)]%
        {hacohen2024similaritiescreatedequalleveraging}
\bibfield{author}{\bibinfo{person}{Uri Hacohen}, \bibinfo{person}{Adi Haviv}, \bibinfo{person}{Shahar Sarfaty}, \bibinfo{person}{Bruria Friedman}, \bibinfo{person}{Niva Elkin-Koren}, \bibinfo{person}{Roi Livni}, {and} \bibinfo{person}{Amit~H Bermano}.} \bibinfo{year}{2024}\natexlab{}.
\newblock \bibinfo{title}{Not All Similarities Are Created Equal: Leveraging Data-Driven Biases to Inform GenAI Copyright Disputes}.
\newblock
\newblock
\showeprint[arxiv]{2403.17691}~[cs.CV]
\urldef\tempurl%
\url{https://arxiv.org/abs/2403.17691}
\showURL{%
\tempurl}


\bibitem[Hagemann et~al\mbox{.}(2018)]%
        {hagemann2018soft}
\bibfield{author}{\bibinfo{person}{Ryan Hagemann}, \bibinfo{person}{Jennifer~Huddleston Skees}, {and} \bibinfo{person}{Adam Thierer}.} \bibinfo{year}{2018}\natexlab{}.
\newblock \showarticletitle{Soft Law for Hard Problems: The Governance of Emerging Technologies in an Uncertain Future}.
\newblock \bibinfo{journal}{\emph{Colorado Technology Law Journal}}  \bibinfo{volume}{17} (\bibinfo{year}{2018}), \bibinfo{pages}{37}.
\newblock


\bibitem[Hartmann et~al\mbox{.}(2023)]%
        {hartmann2023sokmemorizationgeneralpurposelarge}
\bibfield{author}{\bibinfo{person}{Valentin Hartmann}, \bibinfo{person}{Anshuman Suri}, \bibinfo{person}{Vincent Bindschaedler}, \bibinfo{person}{David Evans}, \bibinfo{person}{Shruti Tople}, {and} \bibinfo{person}{Robert West}.} \bibinfo{year}{2023}\natexlab{}.
\newblock \bibinfo{title}{SoK: Memorization in General-Purpose Large Language Models}.
\newblock
\newblock
\showeprint[arxiv]{2310.18362}~[cs.CL]
\urldef\tempurl%
\url{https://arxiv.org/abs/2310.18362}
\showURL{%
\tempurl}


\bibitem[Henderson et~al\mbox{.}(2023)]%
        {henderson_2023_foundation}
\bibfield{author}{\bibinfo{person}{Peter Henderson}, \bibinfo{person}{Xuechen Li}, \bibinfo{person}{Dan Jurafsky}, \bibinfo{person}{Tatsunori Hashimoto}, \bibinfo{person}{Mark~A. Lemley}, {and} \bibinfo{person}{Percy Liang}.} \bibinfo{year}{2023}\natexlab{}.
\newblock \showarticletitle{Foundation Models and Fair Use}.
\newblock \bibinfo{journal}{\emph{Journal of Machine Learning Research}} \bibinfo{volume}{24}, \bibinfo{number}{400} (\bibinfo{year}{2023}), \bibinfo{pages}{1--79}.
\newblock
\urldef\tempurl%
\url{http://jmlr.org/papers/v24/23-0569.html}
\showURL{%
\tempurl}


\bibitem[Huang et~al\mbox{.}(2024)]%
        {huang-etal-2024-demystifying}
\bibfield{author}{\bibinfo{person}{Jing Huang}, \bibinfo{person}{Diyi Yang}, {and} \bibinfo{person}{Christopher Potts}.} \bibinfo{year}{2024}\natexlab{}.
\newblock \showarticletitle{Demystifying Verbatim Memorization in Large Language Models}. In \bibinfo{booktitle}{\emph{Proceedings of the 2024 Conference on Empirical Methods in Natural Language Processing}}, \bibfield{editor}{\bibinfo{person}{Yaser Al-Onaizan}, \bibinfo{person}{Mohit Bansal}, {and} \bibinfo{person}{Yun-Nung Chen}} (Eds.). \bibinfo{publisher}{Association for Computational Linguistics}, \bibinfo{address}{Miami, Florida, USA}, \bibinfo{pages}{10711--10732}.
\newblock
\urldef\tempurl%
\url{https://doi.org/10.18653/v1/2024.emnlp-main.598}
\showDOI{\tempurl}


\bibitem[Ippolito et~al\mbox{.}(2023)]%
        {ippolito-etal-2023-preventing}
\bibfield{author}{\bibinfo{person}{Daphne Ippolito}, \bibinfo{person}{Florian Tramer}, \bibinfo{person}{Milad Nasr}, \bibinfo{person}{Chiyuan Zhang}, \bibinfo{person}{Matthew Jagielski}, \bibinfo{person}{Katherine Lee}, \bibinfo{person}{Christopher Choquette~Choo}, {and} \bibinfo{person}{Nicholas Carlini}.} \bibinfo{year}{2023}\natexlab{}.
\newblock \showarticletitle{Preventing Generation of Verbatim Memorization in Language Models Gives a False Sense of Privacy}. In \bibinfo{booktitle}{\emph{Proceedings of the 16th International Natural Language Generation Conference}}, \bibfield{editor}{\bibinfo{person}{C.~Maria Keet}, \bibinfo{person}{Hung-Yi Lee}, {and} \bibinfo{person}{Sina Zarrie{\ss}}} (Eds.). \bibinfo{publisher}{Association for Computational Linguistics}, \bibinfo{address}{Prague, Czechia}, \bibinfo{pages}{28--53}.
\newblock
\urldef\tempurl%
\url{https://doi.org/10.18653/v1/2023.inlg-main.3}
\showDOI{\tempurl}


\bibitem[Jurafsky and Martin(2024)]%
        {slp_book}
\bibfield{author}{\bibinfo{person}{Daniel Jurafsky} {and} \bibinfo{person}{James~H. Martin}.} \bibinfo{year}{2024}\natexlab{}.
\newblock \bibinfo{booktitle}{\emph{Speech and Language Processing: An Introduction to Natural Language Processing, Computational Linguistics, and Speech Recognition with Language Models} (\bibinfo{edition}{3rd} ed.)}.
\newblock
\urldef\tempurl%
\url{https://web.stanford.edu/~jurafsky/slp3/}
\showURL{%
\tempurl}
\newblock
\shownote{Online manuscript released August 20, 2024}.


\bibitem[Kandpal et~al\mbox{.}(2023)]%
        {pmlr-v202-kandpal23a}
\bibfield{author}{\bibinfo{person}{Nikhil Kandpal}, \bibinfo{person}{Haikang Deng}, \bibinfo{person}{Adam Roberts}, \bibinfo{person}{Eric Wallace}, {and} \bibinfo{person}{Colin Raffel}.} \bibinfo{year}{2023}\natexlab{}.
\newblock \showarticletitle{Large Language Models Struggle to Learn Long-Tail Knowledge}. In \bibinfo{booktitle}{\emph{Proceedings of the 40th International Conference on Machine Learning}} \emph{(\bibinfo{series}{Proceedings of Machine Learning Research}, Vol.~\bibinfo{volume}{202})}, \bibfield{editor}{\bibinfo{person}{Andreas Krause}, \bibinfo{person}{Emma Brunskill}, \bibinfo{person}{Kyunghyun Cho}, \bibinfo{person}{Barbara Engelhardt}, \bibinfo{person}{Sivan Sabato}, {and} \bibinfo{person}{Jonathan Scarlett}} (Eds.). \bibinfo{publisher}{PMLR}, \bibinfo{pages}{15696--15707}.
\newblock
\urldef\tempurl%
\url{https://proceedings.mlr.press/v202/kandpal23a.html}
\showURL{%
\tempurl}


\bibitem[Kandpal et~al\mbox{.}(2022)]%
        {pmlr-v162-kandpal22a}
\bibfield{author}{\bibinfo{person}{Nikhil Kandpal}, \bibinfo{person}{Eric Wallace}, {and} \bibinfo{person}{Colin Raffel}.} \bibinfo{year}{2022}\natexlab{}.
\newblock \showarticletitle{Deduplicating Training Data Mitigates Privacy Risks in Language Models}. In \bibinfo{booktitle}{\emph{Proceedings of the 39th International Conference on Machine Learning}} \emph{(\bibinfo{series}{Proceedings of Machine Learning Research}, Vol.~\bibinfo{volume}{162})}, \bibfield{editor}{\bibinfo{person}{Kamalika Chaudhuri}, \bibinfo{person}{Stefanie Jegelka}, \bibinfo{person}{Le~Song}, \bibinfo{person}{Csaba Szepesvari}, \bibinfo{person}{Gang Niu}, {and} \bibinfo{person}{Sivan Sabato}} (Eds.). \bibinfo{publisher}{PMLR}, \bibinfo{pages}{10697--10707}.
\newblock
\urldef\tempurl%
\url{https://proceedings.mlr.press/v162/kandpal22a.html}
\showURL{%
\tempurl}


\bibitem[Kaplan et~al\mbox{.}(2020)]%
        {kaplan2020scalinglawsneurallanguage}
\bibfield{author}{\bibinfo{person}{Jared Kaplan}, \bibinfo{person}{Sam McCandlish}, \bibinfo{person}{Tom Henighan}, \bibinfo{person}{Tom~B. Brown}, \bibinfo{person}{Benjamin Chess}, \bibinfo{person}{Rewon Child}, \bibinfo{person}{Scott Gray}, \bibinfo{person}{Alec Radford}, \bibinfo{person}{Jeffrey Wu}, {and} \bibinfo{person}{Dario Amodei}.} \bibinfo{year}{2020}\natexlab{}.
\newblock \bibinfo{title}{Scaling Laws for Neural Language Models}.
\newblock
\newblock
\showeprint[arxiv]{2001.08361}~[cs.LG]
\urldef\tempurl%
\url{https://arxiv.org/abs/2001.08361}
\showURL{%
\tempurl}


\bibitem[Kaplow(1992)]%
        {kaplow1992rules}
\bibfield{author}{\bibinfo{person}{Louis Kaplow}.} \bibinfo{year}{1992}\natexlab{}.
\newblock \showarticletitle{Rules Versus Standards: An Economic Analysis}.
\newblock \bibinfo{journal}{\emph{Duke Law Journal}}  \bibinfo{volume}{42} (\bibinfo{year}{1992}), \bibinfo{pages}{557--629}.
\newblock
\urldef\tempurl%
\url{https://scholarship.law.duke.edu/dlj/vol42/iss3/2}
\showURL{%
\tempurl}


\bibitem[Kaplow(2011)]%
        {kaplow2011burden}
\bibfield{author}{\bibinfo{person}{Louis Kaplow}.} \bibinfo{year}{2011}\natexlab{}.
\newblock \showarticletitle{Burden of proof}.
\newblock \bibinfo{journal}{\emph{Yale Law Journal}}  \bibinfo{volume}{121} (\bibinfo{year}{2011}), \bibinfo{pages}{738}.
\newblock


\bibitem[Kirchenbauer et~al\mbox{.}(2024)]%
        {kirchenbauer2024lmd}
\bibfield{author}{\bibinfo{person}{John Kirchenbauer}, \bibinfo{person}{Garrett Honke}, \bibinfo{person}{Gowthami Somepalli}, \bibinfo{person}{Jonas Geiping}, \bibinfo{person}{Katherine Lee}, \bibinfo{person}{Daphne Ippolito}, \bibinfo{person}{Tom Goldstein}, {and} \bibinfo{person}{David Andre}.} \bibinfo{year}{2024}\natexlab{}.
\newblock \showarticletitle{{LMD}3: Language Model Data Density Dependence}. In \bibinfo{booktitle}{\emph{First Conference on Language Modeling}}.
\newblock
\urldef\tempurl%
\url{https://openreview.net/forum?id=eGCw1UVOhk}
\showURL{%
\tempurl}


\bibitem[Lee et~al\mbox{.}(2024)]%
        {lee2024talkinboutaigeneration}
\bibfield{author}{\bibinfo{person}{Katherine Lee}, \bibinfo{person}{A.~Feder Cooper}, {and} \bibinfo{person}{James Grimmelmann}.} \bibinfo{year}{2024}\natexlab{}.
\newblock \bibinfo{title}{Talkin' 'Bout AI Generation: Copyright and the Generative-AI Supply Chain}.
\newblock
\newblock
\showeprint[arxiv]{2309.08133}~[cs.CY]
\urldef\tempurl%
\url{https://arxiv.org/abs/2309.08133}
\showURL{%
\tempurl}


\bibitem[Lee et~al\mbox{.}(2022)]%
        {lee-etal-2022-deduplicating}
\bibfield{author}{\bibinfo{person}{Katherine Lee}, \bibinfo{person}{Daphne Ippolito}, \bibinfo{person}{Andrew Nystrom}, \bibinfo{person}{Chiyuan Zhang}, \bibinfo{person}{Douglas Eck}, \bibinfo{person}{Chris Callison-Burch}, {and} \bibinfo{person}{Nicholas Carlini}.} \bibinfo{year}{2022}\natexlab{}.
\newblock \showarticletitle{Deduplicating Training Data Makes Language Models Better}. In \bibinfo{booktitle}{\emph{Proceedings of the 60th Annual Meeting of the Association for Computational Linguistics (Volume 1: Long Papers)}}, \bibfield{editor}{\bibinfo{person}{Smaranda Muresan}, \bibinfo{person}{Preslav Nakov}, {and} \bibinfo{person}{Aline Villavicencio}} (Eds.). \bibinfo{publisher}{Association for Computational Linguistics}, \bibinfo{address}{Dublin, Ireland}, \bibinfo{pages}{8424--8445}.
\newblock
\urldef\tempurl%
\url{https://doi.org/10.18653/v1/2022.acl-long.577}
\showDOI{\tempurl}


\bibitem[Lemley(2017)]%
        {lemley2017rationalizing}
\bibfield{author}{\bibinfo{person}{Mark~A Lemley}.} \bibinfo{year}{2017}\natexlab{}.
\newblock \showarticletitle{Rationalizing internet safe harbors}.
\newblock In \bibinfo{booktitle}{\emph{Copyright Law}}. \bibinfo{publisher}{Routledge}, \bibinfo{pages}{291--309}.
\newblock


\bibitem[Lemley(2023)]%
        {lemley2023generativeAI}
\bibfield{author}{\bibinfo{person}{Mark~A. Lemley}.} \bibinfo{year}{2023}\natexlab{}.
\newblock \showarticletitle{How Generative AI Turns Copyright Upside Down}.
\newblock  (\bibinfo{date}{July 21} \bibinfo{year}{2023}).
\newblock
\newblock
\shownote{Available at SSRN: \url{https://ssrn.com/abstract=4517702} or \url{http://dx.doi.org/10.2139/ssrn.4517702}}.


\bibitem[Lemley and Casey(2020)]%
        {lemley2020fair}
\bibfield{author}{\bibinfo{person}{Mark~A. Lemley} {and} \bibinfo{person}{Bryan Casey}.} \bibinfo{year}{2020}\natexlab{}.
\newblock \showarticletitle{Fair Learning}.
\newblock  (\bibinfo{date}{January 30} \bibinfo{year}{2020}).
\newblock
\newblock
\shownote{Available at SSRN: \url{https://ssrn.com/abstract=3528447} or \url{http://dx.doi.org/10.2139/ssrn.3528447}}.


\bibitem[Lesci et~al\mbox{.}(2024)]%
        {lesci-etal-2024-causal}
\bibfield{author}{\bibinfo{person}{Pietro Lesci}, \bibinfo{person}{Clara Meister}, \bibinfo{person}{Thomas Hofmann}, \bibinfo{person}{Andreas Vlachos}, {and} \bibinfo{person}{Tiago Pimentel}.} \bibinfo{year}{2024}\natexlab{}.
\newblock \showarticletitle{Causal Estimation of Memorisation Profiles}. In \bibinfo{booktitle}{\emph{Proceedings of the 62nd Annual Meeting of the Association for Computational Linguistics (Volume 1: Long Papers)}}, \bibfield{editor}{\bibinfo{person}{Lun-Wei Ku}, \bibinfo{person}{Andre Martins}, {and} \bibinfo{person}{Vivek Srikumar}} (Eds.). \bibinfo{publisher}{Association for Computational Linguistics}, \bibinfo{address}{Bangkok, Thailand}, \bibinfo{pages}{15616--15635}.
\newblock
\urldef\tempurl%
\url{https://aclanthology.org/2024.acl-long.834}
\showURL{%
\tempurl}


\bibitem[Liesenfeld and Dingemanse(2024)]%
        {liesenfeld_rethinking_2024}
\bibfield{author}{\bibinfo{person}{Andreas Liesenfeld} {and} \bibinfo{person}{Mark Dingemanse}.} \bibinfo{year}{2024}\natexlab{}.
\newblock \showarticletitle{Rethinking open source generative AI: open-washing and the EU AI Act}. In \bibinfo{booktitle}{\emph{Proceedings of the 2024 ACM Conference on Fairness, Accountability, and Transparency}} (Rio de Janeiro, Brazil) \emph{(\bibinfo{series}{FAccT '24})}. \bibinfo{publisher}{Association for Computing Machinery}, \bibinfo{address}{New York, NY, USA}, \bibinfo{pages}{1774–1787}.
\newblock
\showISBNx{9798400704505}
\urldef\tempurl%
\url{https://doi.org/10.1145/3630106.3659005}
\showDOI{\tempurl}


\bibitem[Lim(2021)]%
        {lim2021substantial}
\bibfield{author}{\bibinfo{person}{Daryl Lim}.} \bibinfo{year}{2021}\natexlab{}.
\newblock \showarticletitle{Saving Substantial Similarity}.
\newblock \bibinfo{journal}{\emph{Florida Law Review}}  \bibinfo{volume}{73} (\bibinfo{date}{May} \bibinfo{year}{2021}), \bibinfo{pages}{591}.
\newblock


\bibitem[Longpre et~al\mbox{.}(2024)]%
        {longpre-etal-2024-pretrainers}
\bibfield{author}{\bibinfo{person}{Shayne Longpre}, \bibinfo{person}{Gregory Yauney}, \bibinfo{person}{Emily Reif}, \bibinfo{person}{Katherine Lee}, \bibinfo{person}{Adam Roberts}, \bibinfo{person}{Barret Zoph}, \bibinfo{person}{Denny Zhou}, \bibinfo{person}{Jason Wei}, \bibinfo{person}{Kevin Robinson}, \bibinfo{person}{David Mimno}, {and} \bibinfo{person}{Daphne Ippolito}.} \bibinfo{year}{2024}\natexlab{}.
\newblock \showarticletitle{A Pretrainer`s Guide to Training Data: Measuring the Effects of Data Age, Domain Coverage, Quality, {\&} Toxicity}. In \bibinfo{booktitle}{\emph{Proceedings of the 2024 Conference of the North American Chapter of the Association for Computational Linguistics: Human Language Technologies (Volume 1: Long Papers)}}, \bibfield{editor}{\bibinfo{person}{Kevin Duh}, \bibinfo{person}{Helena Gomez}, {and} \bibinfo{person}{Steven Bethard}} (Eds.). \bibinfo{publisher}{Association for Computational Linguistics}, \bibinfo{address}{Mexico City, Mexico}, \bibinfo{pages}{3245--3276}.
\newblock
\urldef\tempurl%
\url{https://doi.org/10.18653/v1/2024.naacl-long.179}
\showDOI{\tempurl}


\bibitem[Marchant(2020)]%
        {marchant2020governance}
\bibfield{author}{\bibinfo{person}{Gary~E. Marchant}.} \bibinfo{year}{2020}\natexlab{}.
\newblock \showarticletitle{Governance of Emerging Technologies as a Wicked Problem}.
\newblock \bibinfo{journal}{\emph{Vanderbilt Law Review}}  \bibinfo{volume}{73} (\bibinfo{year}{2020}), \bibinfo{pages}{1861}.
\newblock
\urldef\tempurl%
\url{https://scholarship.law.vanderbilt.edu/vlr/vol73/iss6/8}
\showURL{%
\tempurl}


\bibitem[Meng et~al\mbox{.}(2022)]%
        {meng_locating_2022}
\bibfield{author}{\bibinfo{person}{Kevin Meng}, \bibinfo{person}{David Bau}, \bibinfo{person}{Alex Andonian}, {and} \bibinfo{person}{Yonatan Belinkov}.} \bibinfo{year}{2022}\natexlab{}.
\newblock \showarticletitle{Locating and Editing Factual Associations in GPT}. In \bibinfo{booktitle}{\emph{Advances in Neural Information Processing Systems}}, \bibfield{editor}{\bibinfo{person}{S.~Koyejo}, \bibinfo{person}{S.~Mohamed}, \bibinfo{person}{A.~Agarwal}, \bibinfo{person}{D.~Belgrave}, \bibinfo{person}{K.~Cho}, {and} \bibinfo{person}{A.~Oh}} (Eds.), Vol.~\bibinfo{volume}{35}. \bibinfo{publisher}{Curran Associates, Inc.}, \bibinfo{pages}{17359--17372}.
\newblock
\urldef\tempurl%
\url{https://proceedings.neurips.cc/paper_files/paper/2022/file/6f1d43d5a82a37e89b0665b33bf3a182-Paper-Conference.pdf}
\showURL{%
\tempurl}


\bibitem[Merrill et~al\mbox{.}(2024)]%
        {merrill-etal-2024-evaluating}
\bibfield{author}{\bibinfo{person}{William Merrill}, \bibinfo{person}{Noah~A. Smith}, {and} \bibinfo{person}{Yanai Elazar}.} \bibinfo{year}{2024}\natexlab{}.
\newblock \showarticletitle{Evaluating $n$-Gram Novelty of Language Models Using Rusty-{DAWG}}. In \bibinfo{booktitle}{\emph{Proceedings of the 2024 Conference on Empirical Methods in Natural Language Processing}}, \bibfield{editor}{\bibinfo{person}{Yaser Al-Onaizan}, \bibinfo{person}{Mohit Bansal}, {and} \bibinfo{person}{Yun-Nung Chen}} (Eds.). \bibinfo{publisher}{Association for Computational Linguistics}, \bibinfo{address}{Miami, Florida, USA}, \bibinfo{pages}{14459--14473}.
\newblock
\urldef\tempurl%
\url{https://doi.org/10.18653/v1/2024.emnlp-main.800}
\showDOI{\tempurl}


\bibitem[Mitchell et~al\mbox{.}(2019)]%
        {mitchell_2019_model}
\bibfield{author}{\bibinfo{person}{Margaret Mitchell}, \bibinfo{person}{Simone Wu}, \bibinfo{person}{Andrew Zaldivar}, \bibinfo{person}{Parker Barnes}, \bibinfo{person}{Lucy Vasserman}, \bibinfo{person}{Ben Hutchinson}, \bibinfo{person}{Elena Spitzer}, \bibinfo{person}{Inioluwa~Deborah Raji}, {and} \bibinfo{person}{Timnit Gebru}.} \bibinfo{year}{2019}\natexlab{}.
\newblock \showarticletitle{Model Cards for Model Reporting}. In \bibinfo{booktitle}{\emph{Proceedings of the Conference on Fairness, Accountability, and Transparency}} (Atlanta, GA, USA) \emph{(\bibinfo{series}{FAT* '19})}. \bibinfo{publisher}{Association for Computing Machinery}, \bibinfo{address}{New York, NY, USA}, \bibinfo{pages}{220–229}.
\newblock
\showISBNx{9781450361255}
\urldef\tempurl%
\url{https://doi.org/10.1145/3287560.3287596}
\showDOI{\tempurl}


\bibitem[{OpenAI}(2025)]%
        {openai_journalism}
\bibfield{author}{\bibinfo{person}{{OpenAI}}.} \bibinfo{year}{2025}\natexlab{}.
\newblock \bibinfo{title}{{OpenAI and Journalism}}.
\newblock
\newblock
\urldef\tempurl%
\url{https://openai.com/index/openai-and-journalism/}
\showURL{%
\tempurl}
\newblock
\shownote{Accessed: 2025-01-21}.


\bibitem[Sag(2023)]%
        {sag2023copyright}
\bibfield{author}{\bibinfo{person}{Matthew Sag}.} \bibinfo{year}{2023}\natexlab{}.
\newblock \showarticletitle{Copyright Safety for Generative AI}.
\newblock \bibinfo{journal}{\emph{Houston Law Review}} \bibinfo{volume}{61}, \bibinfo{number}{2} (\bibinfo{year}{2023}).
\newblock
\urldef\tempurl%
\url{https://doi.org/10.2139/ssrn.4438593}
\showDOI{\tempurl}


\bibitem[Schauer(2009)]%
        {schauer_thinking_2009}
\bibfield{author}{\bibinfo{person}{Frederick Schauer}.} \bibinfo{year}{2009}\natexlab{}.
\newblock \bibinfo{booktitle}{\emph{Thinking Like a Lawyer: A New Introduction to Legal Reasoning}}.
\newblock \bibinfo{publisher}{Harvard University Press}.
\newblock
\showISBNx{9780674032705}
\urldef\tempurl%
\url{http://www.jstor.org/stable/j.ctvjk2x3k}
\showURL{%
\tempurl}


\bibitem[Scheffler et~al\mbox{.}(2022)]%
        {scheffler_2022_formalizing}
\bibfield{author}{\bibinfo{person}{Sarah Scheffler}, \bibinfo{person}{Eran Tromer}, {and} \bibinfo{person}{Mayank Varia}.} \bibinfo{year}{2022}\natexlab{}.
\newblock \showarticletitle{Formalizing Human Ingenuity: A Quantitative Framework for Copyright Law's Substantial Similarity}. In \bibinfo{booktitle}{\emph{Proceedings of the 2022 Symposium on Computer Science and Law}} (Washington DC, USA) \emph{(\bibinfo{series}{CSLAW '22})}. \bibinfo{publisher}{Association for Computing Machinery}, \bibinfo{address}{New York, NY, USA}, \bibinfo{pages}{37–49}.
\newblock
\showISBNx{9781450392341}
\urldef\tempurl%
\url{https://doi.org/10.1145/3511265.3550444}
\showDOI{\tempurl}


\bibitem[Schwarzschild et~al\mbox{.}(2024)]%
        {schwarzschild2024rethinkingllmmemorizationlens}
\bibfield{author}{\bibinfo{person}{Avi Schwarzschild}, \bibinfo{person}{Zhili Feng}, \bibinfo{person}{Pratyush Maini}, \bibinfo{person}{Zachary~C. Lipton}, {and} \bibinfo{person}{J.~Zico Kolter}.} \bibinfo{year}{2024}\natexlab{}.
\newblock \bibinfo{title}{Rethinking LLM Memorization through the Lens of Adversarial Compression}.
\newblock
\newblock
\showeprint[arxiv]{2404.15146}~[cs.LG]
\urldef\tempurl%
\url{https://arxiv.org/abs/2404.15146}
\showURL{%
\tempurl}


\bibitem[Selbst(2021)]%
        {selbst2021institutional}
\bibfield{author}{\bibinfo{person}{Andrew~D. Selbst}.} \bibinfo{year}{2021}\natexlab{}.
\newblock \showarticletitle{An Institutional View of Algorithmic Impact Assessments}.
\newblock \bibinfo{journal}{\emph{Harvard Journal of Law \& Technology}}  \bibinfo{volume}{35} (\bibinfo{year}{2021}), \bibinfo{pages}{117}.
\newblock
\urldef\tempurl%
\url{https://ssrn.com/abstract=3867634}
\showURL{%
\tempurl}


\bibitem[Selbst et~al\mbox{.}(2024)]%
        {selbst2024deconstructing}
\bibfield{author}{\bibinfo{person}{Andrew~D. Selbst}, \bibinfo{person}{Suresh Venkatasubramanian}, {and} \bibinfo{person}{I.~Elizabeth Kumar}.} \bibinfo{year}{2024}\natexlab{}.
\newblock \showarticletitle{Deconstructing Design Decisions: Why Courts Must Interrogate Machine Learning and Other Technologies}.
\newblock \bibinfo{journal}{\emph{Ohio State Law Journal}}  \bibinfo{volume}{85} (\bibinfo{date}{September 7} \bibinfo{year}{2024}), \bibinfo{pages}{415}.
\newblock
\urldef\tempurl%
\url{https://ssrn.com/abstract=4564304}
\showURL{%
\tempurl}
\newblock
\shownote{UCLA School of Law, Public Law Research Paper No. 23-22}.


\bibitem[Shackelford et~al\mbox{.}(2015)]%
        {shackelford2015global}
\bibfield{author}{\bibinfo{person}{Scott~J. Shackelford}, \bibinfo{person}{Andrew~A. Proia}, \bibinfo{person}{Brenton Martell}, {and} \bibinfo{person}{Amanda~N. Craig}.} \bibinfo{year}{2015}\natexlab{}.
\newblock \showarticletitle{Toward a Global Cybersecurity Standard of Care: Exploring the Implications of the 2014 NIST Cybersecurity Framework on Shaping Reasonable National and International Cybersecurity Practices}.
\newblock \bibinfo{journal}{\emph{Texas International Law Journal}} \bibinfo{volume}{50}, \bibinfo{number}{2} (\bibinfo{year}{2015}), \bibinfo{pages}{305--}.
\newblock
\newblock
\shownote{Spring-Summer}.


\bibitem[Shi et~al\mbox{.}(2024)]%
        {shi2024detecting}
\bibfield{author}{\bibinfo{person}{Weijia Shi}, \bibinfo{person}{Anirudh Ajith}, \bibinfo{person}{Mengzhou Xia}, \bibinfo{person}{Yangsibo Huang}, \bibinfo{person}{Daogao Liu}, \bibinfo{person}{Terra Blevins}, \bibinfo{person}{Danqi Chen}, {and} \bibinfo{person}{Luke Zettlemoyer}.} \bibinfo{year}{2024}\natexlab{}.
\newblock \showarticletitle{Detecting Pretraining Data from Large Language Models}. In \bibinfo{booktitle}{\emph{The Twelfth International Conference on Learning Representations}}.
\newblock
\urldef\tempurl%
\url{https://openreview.net/forum?id=zWqr3MQuNs}
\showURL{%
\tempurl}


\bibitem[Shokri et~al\mbox{.}(2017)]%
        {DBLP:conf/sp/ShokriSSS17}
\bibfield{author}{\bibinfo{person}{Reza Shokri}, \bibinfo{person}{Marco Stronati}, \bibinfo{person}{Congzheng Song}, {and} \bibinfo{person}{Vitaly Shmatikov}.} \bibinfo{year}{2017}\natexlab{}.
\newblock \showarticletitle{Membership Inference Attacks Against Machine Learning Models}. In \bibinfo{booktitle}{\emph{2017 {IEEE} Symposium on Security and Privacy, {SP} 2017, San Jose, CA, USA, May 22-26, 2017}}. \bibinfo{publisher}{{IEEE} Computer Society}, \bibinfo{pages}{3--18}.
\newblock
\urldef\tempurl%
\url{https://doi.org/10.1109/SP.2017.41}
\showDOI{\tempurl}


\bibitem[Sobel(2017)]%
        {sobel2017fairuse}
\bibfield{author}{\bibinfo{person}{Benjamin Sobel}.} \bibinfo{year}{2017}\natexlab{}.
\newblock \showarticletitle{Artificial Intelligence's Fair Use Crisis}.
\newblock \bibinfo{journal}{\emph{Columbia Journal of Law \& the Arts}}  \bibinfo{volume}{41} (\bibinfo{date}{September 4} \bibinfo{year}{2017}), \bibinfo{pages}{45}.
\newblock
\newblock
\shownote{Available at SSRN: \url{https://ssrn.com/abstract=3032076}}.


\bibitem[Somepalli et~al\mbox{.}(2023)]%
        {DBLP:conf/cvpr/SomepalliSGGG23}
\bibfield{author}{\bibinfo{person}{Gowthami Somepalli}, \bibinfo{person}{Vasu Singla}, \bibinfo{person}{Micah Goldblum}, \bibinfo{person}{Jonas Geiping}, {and} \bibinfo{person}{Tom Goldstein}.} \bibinfo{year}{2023}\natexlab{}.
\newblock \showarticletitle{Diffusion Art or Digital Forgery? Investigating Data Replication in Diffusion Models}. In \bibinfo{booktitle}{\emph{{IEEE/CVF} Conference on Computer Vision and Pattern Recognition, {CVPR} 2023, Vancouver, BC, Canada, June 17-24, 2023}}. \bibinfo{publisher}{{IEEE}}, \bibinfo{pages}{6048--6058}.
\newblock
\urldef\tempurl%
\url{https://doi.org/10.1109/CVPR52729.2023.00586}
\showDOI{\tempurl}


\bibitem[Tirumala et~al\mbox{.}(2022)]%
        {tirumala-memorization-2022}
\bibfield{author}{\bibinfo{person}{Kushal Tirumala}, \bibinfo{person}{Aram Markosyan}, \bibinfo{person}{Luke Zettlemoyer}, {and} \bibinfo{person}{Armen Aghajanyan}.} \bibinfo{year}{2022}\natexlab{}.
\newblock \showarticletitle{Memorization Without Overfitting: Analyzing the Training Dynamics of Large Language Models}. In \bibinfo{booktitle}{\emph{Advances in Neural Information Processing Systems}}, \bibfield{editor}{\bibinfo{person}{S.~Koyejo}, \bibinfo{person}{S.~Mohamed}, \bibinfo{person}{A.~Agarwal}, \bibinfo{person}{D.~Belgrave}, \bibinfo{person}{K.~Cho}, {and} \bibinfo{person}{A.~Oh}} (Eds.), Vol.~\bibinfo{volume}{35}. \bibinfo{publisher}{Curran Associates, Inc.}, \bibinfo{pages}{38274--38290}.
\newblock
\urldef\tempurl%
\url{https://proceedings.neurips.cc/paper_files/paper/2022/file/fa0509f4dab6807e2cb465715bf2d249-Paper-Conference.pdf}
\showURL{%
\tempurl}


\bibitem[Vaswani et~al\mbox{.}(2017)]%
        {transformers}
\bibfield{author}{\bibinfo{person}{Ashish Vaswani}, \bibinfo{person}{Noam Shazeer}, \bibinfo{person}{Niki Parmar}, \bibinfo{person}{Jakob Uszkoreit}, \bibinfo{person}{Llion Jones}, \bibinfo{person}{Aidan~N. Gomez}, \bibinfo{person}{\L{}ukasz Kaiser}, {and} \bibinfo{person}{Illia Polosukhin}.} \bibinfo{year}{2017}\natexlab{}.
\newblock \showarticletitle{Attention is all you need}. In \bibinfo{booktitle}{\emph{Proceedings of the 31st International Conference on Neural Information Processing Systems}} (Long Beach, California, USA) \emph{(\bibinfo{series}{NIPS'17})}. \bibinfo{publisher}{Curran Associates Inc.}, \bibinfo{address}{Red Hook, NY, USA}, \bibinfo{pages}{6000–6010}.
\newblock
\showISBNx{9781510860964}


\bibitem[Vyas et~al\mbox{.}(2023)]%
        {pmlr-v202-vyas23b}
\bibfield{author}{\bibinfo{person}{Nikhil Vyas}, \bibinfo{person}{Sham~M. Kakade}, {and} \bibinfo{person}{Boaz Barak}.} \bibinfo{year}{2023}\natexlab{}.
\newblock \showarticletitle{On Provable Copyright Protection for Generative Models}. In \bibinfo{booktitle}{\emph{Proceedings of the 40th International Conference on Machine Learning}} \emph{(\bibinfo{series}{Proceedings of Machine Learning Research}, Vol.~\bibinfo{volume}{202})}, \bibfield{editor}{\bibinfo{person}{Andreas Krause}, \bibinfo{person}{Emma Brunskill}, \bibinfo{person}{Kyunghyun Cho}, \bibinfo{person}{Barbara Engelhardt}, \bibinfo{person}{Sivan Sabato}, {and} \bibinfo{person}{Jonathan Scarlett}} (Eds.). \bibinfo{publisher}{PMLR}, \bibinfo{pages}{35277--35299}.
\newblock
\urldef\tempurl%
\url{https://proceedings.mlr.press/v202/vyas23b.html}
\showURL{%
\tempurl}


\bibitem[Wagner(2000)]%
        {wagner2000triumph}
\bibfield{author}{\bibinfo{person}{Wendy~E Wagner}.} \bibinfo{year}{2000}\natexlab{}.
\newblock \showarticletitle{The triumph of technology-based standards}.
\newblock \bibinfo{journal}{\emph{U. Ill. L. Rev.}} (\bibinfo{year}{2000}), \bibinfo{pages}{83}.
\newblock


\bibitem[Wallace et~al\mbox{.}(2019)]%
        {wallace-etal-2019-universal}
\bibfield{author}{\bibinfo{person}{Eric Wallace}, \bibinfo{person}{Shi Feng}, \bibinfo{person}{Nikhil Kandpal}, \bibinfo{person}{Matt Gardner}, {and} \bibinfo{person}{Sameer Singh}.} \bibinfo{year}{2019}\natexlab{}.
\newblock \showarticletitle{Universal Adversarial Triggers for Attacking and Analyzing {NLP}}. In \bibinfo{booktitle}{\emph{Proceedings of the 2019 Conference on Empirical Methods in Natural Language Processing and the 9th International Joint Conference on Natural Language Processing (EMNLP-IJCNLP)}}, \bibfield{editor}{\bibinfo{person}{Kentaro Inui}, \bibinfo{person}{Jing Jiang}, \bibinfo{person}{Vincent Ng}, {and} \bibinfo{person}{Xiaojun Wan}} (Eds.). \bibinfo{publisher}{Association for Computational Linguistics}, \bibinfo{address}{Hong Kong, China}, \bibinfo{pages}{2153--2162}.
\newblock
\urldef\tempurl%
\url{https://doi.org/10.18653/v1/D19-1221}
\showDOI{\tempurl}


\bibitem[Wei et~al\mbox{.}(2024a)]%
        {wei2024evaluating}
\bibfield{author}{\bibinfo{person}{Boyi Wei}, \bibinfo{person}{Weijia Shi}, \bibinfo{person}{Yangsibo Huang}, \bibinfo{person}{Noah~A. Smith}, \bibinfo{person}{Chiyuan Zhang}, \bibinfo{person}{Luke Zettlemoyer}, \bibinfo{person}{Kai Li}, {and} \bibinfo{person}{Peter Henderson}.} \bibinfo{year}{2024}\natexlab{a}.
\newblock \showarticletitle{Evaluating Copyright Takedown Methods for Language Models}. In \bibinfo{booktitle}{\emph{The Thirty-eight Conference on Neural Information Processing Systems Datasets and Benchmarks Track}}.
\newblock
\urldef\tempurl%
\url{https://openreview.net/forum?id=ar8aRMrmod}
\showURL{%
\tempurl}


\bibitem[Wei et~al\mbox{.}(2024b)]%
        {wei-etal-2024-proving}
\bibfield{author}{\bibinfo{person}{Johnny Wei}, \bibinfo{person}{Ryan Wang}, {and} \bibinfo{person}{Robin Jia}.} \bibinfo{year}{2024}\natexlab{b}.
\newblock \showarticletitle{Proving membership in {LLM} pretraining data via data watermarks}. In \bibinfo{booktitle}{\emph{Findings of the Association for Computational Linguistics: ACL 2024}}, \bibfield{editor}{\bibinfo{person}{Lun-Wei Ku}, \bibinfo{person}{Andre Martins}, {and} \bibinfo{person}{Vivek Srikumar}} (Eds.). \bibinfo{publisher}{Association for Computational Linguistics}, \bibinfo{address}{Bangkok, Thailand}, \bibinfo{pages}{13306--13320}.
\newblock
\urldef\tempurl%
\url{https://doi.org/10.18653/v1/2024.findings-acl.788}
\showDOI{\tempurl}


\bibitem[Yauney et~al\mbox{.}(2023)]%
        {yauney-etal-2023-data}
\bibfield{author}{\bibinfo{person}{Gregory Yauney}, \bibinfo{person}{Emily Reif}, {and} \bibinfo{person}{David Mimno}.} \bibinfo{year}{2023}\natexlab{}.
\newblock \showarticletitle{Data Similarity is Not Enough to Explain Language Model Performance}. In \bibinfo{booktitle}{\emph{Proceedings of the 2023 Conference on Empirical Methods in Natural Language Processing}}, \bibfield{editor}{\bibinfo{person}{Houda Bouamor}, \bibinfo{person}{Juan Pino}, {and} \bibinfo{person}{Kalika Bali}} (Eds.). \bibinfo{publisher}{Association for Computational Linguistics}, \bibinfo{address}{Singapore}, \bibinfo{pages}{11295--11304}.
\newblock
\urldef\tempurl%
\url{https://doi.org/10.18653/v1/2023.emnlp-main.695}
\showDOI{\tempurl}


\bibitem[Yew(2024)]%
        {yew_break_2024}
\bibfield{author}{\bibinfo{person}{Rui-Jie Yew}.} \bibinfo{year}{2024}\natexlab{}.
\newblock \showarticletitle{Break It 'Til You Make It: An Exploration of the Ramifications of Copyright Liability Under a Pre-training Paradigm of AI Development}. In \bibinfo{booktitle}{\emph{Proceedings of the Symposium on Computer Science and Law}} (Boston, MA, USA) \emph{(\bibinfo{series}{CSLAW '24})}. \bibinfo{publisher}{Association for Computing Machinery}, \bibinfo{address}{New York, NY, USA}, \bibinfo{pages}{64–72}.
\newblock
\showISBNx{9798400703331}
\urldef\tempurl%
\url{https://doi.org/10.1145/3614407.3643707}
\showDOI{\tempurl}


\bibitem[Young et~al\mbox{.}(2022)]%
        {young_confronting_2022}
\bibfield{author}{\bibinfo{person}{Meg Young}, \bibinfo{person}{Michael Katell}, {and} \bibinfo{person}{P.M. Krafft}.} \bibinfo{year}{2022}\natexlab{}.
\newblock \showarticletitle{Confronting Power and Corporate Capture at the FAccT Conference}. In \bibinfo{booktitle}{\emph{Proceedings of the 2022 ACM Conference on Fairness, Accountability, and Transparency}} (Seoul, Republic of Korea) \emph{(\bibinfo{series}{FAccT '22})}. \bibinfo{publisher}{Association for Computing Machinery}, \bibinfo{address}{New York, NY, USA}, \bibinfo{pages}{1375–1386}.
\newblock
\showISBNx{9781450393522}
\urldef\tempurl%
\url{https://doi.org/10.1145/3531146.3533194}
\showDOI{\tempurl}


\bibitem[\y\y(1787)]%
        {us_const_art1_sec8_cl8}
\y\y \bibinfo{year}{1787}\natexlab{}.
\newblock \bibinfo{title}{U.S. Constitution, Article I, Section 8, Clause 8}.
\newblock
\newblock
\newblock
\shownote{“To promote the Progress of Science and useful Arts, by securing for limited Times to Authors and Inventors the exclusive Right to their respective Writings and Discoveries.”}.


\bibitem[\y\y(2024a)]%
        {usc_17_106}
\y\y \bibinfo{year}{2024}\natexlab{a}.
\newblock \bibinfo{title}{17 U.S.C. § 106: Exclusive Rights in Copyrighted Works}.
\newblock \bibinfo{howpublished}{United States Code}.
\newblock
\urldef\tempurl%
\url{https://www.law.cornell.edu/uscode/text/17/106}
\showURL{%
\tempurl}
\newblock
\shownote{This section grants copyright owners exclusive rights to reproduce, prepare derivative works, distribute copies, perform publicly, display publicly, and, in the case of sound recordings, perform publicly by means of digital audio transmission.}.


\bibitem[\y\y(2024b)]%
        {usc_17_107}
\y\y \bibinfo{year}{2024}\natexlab{b}.
\newblock \bibinfo{title}{17 U.S.C. § 107: Limitations on Exclusive Rights—Fair Use}.
\newblock \bibinfo{howpublished}{United States Code}.
\newblock
\urldef\tempurl%
\url{https://www.law.cornell.edu/uscode/text/17/107}
\showURL{%
\tempurl}
\newblock
\shownote{This section outlines the fair use doctrine, permitting certain uses of copyrighted works for purposes such as criticism, comment, news reporting, teaching, scholarship, or research. It specifies four factors to consider in determining fair use: (1) the purpose and character of the use; (2) the nature of the copyrighted work; (3) the amount and substantiality of the portion used; and (4) the effect of the use upon the potential market for or value of the copyrighted work.}.


\bibitem[\z\z\emph{Arnstein v. Porter}(1946)]%
        {arnstein1946}
\bibfield{author}{\bibinfo{person}{\z\z\emph{Arnstein v. Porter}}.} \bibinfo{year}{1946}\natexlab{}.
\newblock \bibinfo{howpublished}{154 F.2d 464, 468 (2d Cir. 1946)}.
\newblock


\bibitem[\z\z\emph{Campbell v. Acuff-Rose Music, Inc.}(1994)]%
        {campbell1994}
\bibfield{author}{\bibinfo{person}{\z\z\emph{Campbell v. Acuff-Rose Music, Inc.}}} \bibinfo{year}{1994}\natexlab{}.
\newblock \bibinfo{howpublished}{510 U.S. 569 (1994)}.
\newblock


\bibitem[\z\z\emph{Computer Assocs. Int'l, Inc. v. Altai, Inc.}(1992)]%
        {computerassociates1992}
\bibfield{author}{\bibinfo{person}{\z\z\emph{Computer Assocs. Int'l, Inc. v. Altai, Inc.}}} \bibinfo{year}{1992}\natexlab{}.
\newblock \bibinfo{howpublished}{982 F.2d 693 (2d Cir. 1992)}.
\newblock


\bibitem[\z\z\emph{Feist Publ'ns, Inc. v. Rural Tel. Serv. Co.}(1991)]%
        {feist1991}
\bibfield{author}{\bibinfo{person}{\z\z\emph{Feist Publ'ns, Inc. v. Rural Tel. Serv. Co.}}} \bibinfo{year}{1991}\natexlab{}.
\newblock \bibinfo{howpublished}{499 U.S. 340 (1991)}.
\newblock


\bibitem[\z\z\emph{Field v. Google Inc.}(2006)]%
        {field2006}
\bibfield{author}{\bibinfo{person}{\z\z\emph{Field v. Google Inc.}}} \bibinfo{year}{2006}\natexlab{}.
\newblock \bibinfo{howpublished}{412 F. Supp. 2d 1106 (D. Nev. 2006)}.
\newblock


\bibitem[\z\z\emph{Harper \& Row, Publ'rs, Inc. v. Nation Enters.}(1985)]%
        {harper1985}
\bibfield{author}{\bibinfo{person}{\z\z\emph{Harper \& Row, Publ'rs, Inc. v. Nation Enters.}}} \bibinfo{year}{1985}\natexlab{}.
\newblock \bibinfo{howpublished}{471 U.S. 539 (1985)}.
\newblock


\bibitem[\z\z\emph{Kadrey v. Meta Platforms, Inc.}(2023)]%
        {kadrey_v_meta}
\bibfield{author}{\bibinfo{person}{\z\z\emph{Kadrey v. Meta Platforms, Inc.}}} \bibinfo{year}{2023}\natexlab{}.
\newblock \bibinfo{howpublished}{3:23-cv-03417}.
\newblock


\bibitem[\z\z\emph{Perfect 10, Inc. v. Amazon.com, Inc.}(2007)]%
        {perfect10_2007}
\bibfield{author}{\bibinfo{person}{\z\z\emph{Perfect 10, Inc. v. Amazon.com, Inc.}}} \bibinfo{year}{2007}\natexlab{}.
\newblock \bibinfo{howpublished}{508 F.3d 1146 (9th Cir. 2007)}.
\newblock


\bibitem[\z\z\emph{The New York Times Co. v. Microsoft Corp.}(2023)]%
        {nytimes_v_microsoft}
\bibfield{author}{\bibinfo{person}{\z\z\emph{The New York Times Co. v. Microsoft Corp.}}} \bibinfo{year}{2023}\natexlab{}.
\newblock \bibinfo{howpublished}{No. 1:2023cv11195, 17 U.S.C. § 501 Copyright Infringement (S.D.N.Y. filed Dec. 27, 2023)}.
\newblock


\end{thebibliography}

%%
%% If your work has an appendix, this is the place to put it.
\newpage
\appendix
\section{Additional results on upweighting}
\label{appendix:upweighting}
\begin{figure}[h]
\centering
\includegraphics[width=0.65\linewidth]{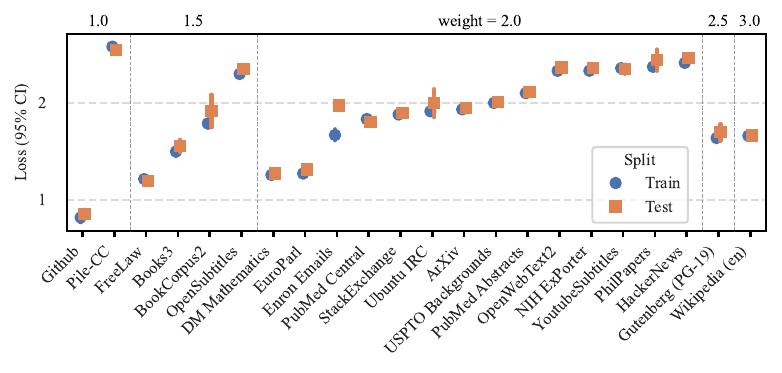}
\includegraphics[width=0.65\linewidth]{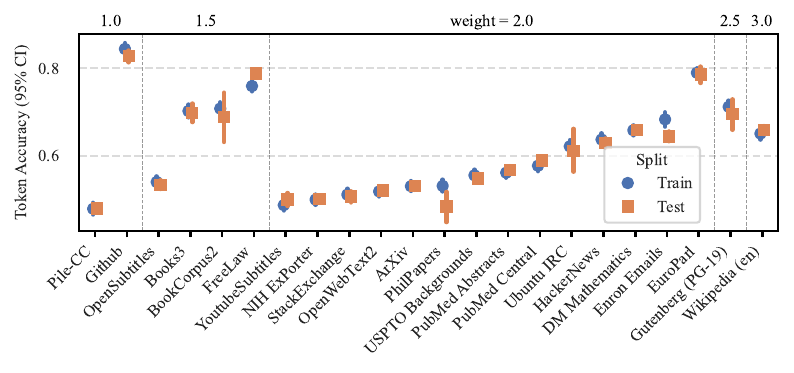}
\includegraphics[width=0.65\linewidth]{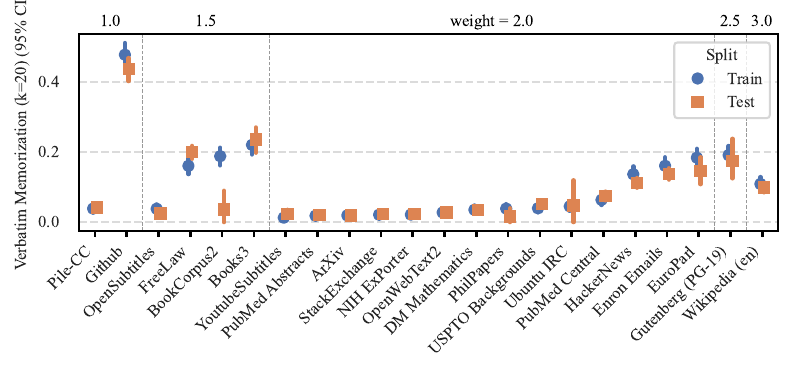}
\caption{(Top) Mean cross-entropy loss, (middle) token-level accuracy, and (bottom) verbatim memorization for the Pythia 6.9B model across training (blue) and test (orange) sets, grouped by dataset component. Error bars denote 95\% CI. These metrics are described in \S\ref{sec:operationalizing}. The difference in the metric between train and test represents the causal effect of upweighting a single document. Differences are grouped by upweight value, assigned by the creators of the Pile. Similar to MINK\%, these metrics vary little across the train and test splits indicating that upweighting a single document has little effect on memorization.}
\label{fig:all_metrics}
\end{figure}
\newpage
\begin{table}[h]
    \centering
\begin{tabular}{lccccccc}
Dataset Name & $\alpha$ & $\beta_1$ & $\alpha$ (95\% CI) & $\beta_1$ (95\% CI) & $r$-squared & Train Mean & Test Mean \\
\midrule
Pile-CC & 6.946 & 0.047 & [6.880, 7.012] & [-0.046, 0.140] & 0.000 & 6.993 & 6.946 \\
PubMed Central & 5.551 & 0.020 & [5.477, 5.626] & [-0.085, 0.125] & 0.000 & 5.572 & 5.551 \\
Books3 & 5.338 & -0.104 & [5.133, 5.543] & [-0.336, 0.127] & 0.001 & 5.233 & 5.338 \\
OpenWebText2 & 6.559 & -0.031 & [6.499, 6.619] & [-0.116, 0.054] & 0.000 & 6.528 & 6.559 \\
ArXiv & 5.899 & -0.047 & [5.844, 5.953] & [-0.124, 0.029] & 0.001 & 5.851 & 5.899 \\
Github & 3.358 & -0.173 & [3.225, 3.491] & [-0.362, 0.016] & 0.002 & 3.185 & 3.358 \\
FreeLaw & 4.734 & 0.048 & [4.660, 4.808] & [-0.058, 0.153] & 0.000 & 4.781 & 4.734 \\
StackExchange & 5.785 & -0.047 & [5.725, 5.845] & [-0.132, 0.038] & 0.001 & 5.738 & 5.785 \\
USPTO Backgrounds & 5.803 & -0.034 & [5.735, 5.871] & [-0.130, 0.062] & 0.000 & 5.769 & 5.803 \\
PubMed Abstracts & 6.143 & -0.039 & [6.091, 6.195] & [-0.112, 0.035] & 0.001 & 6.104 & 6.143 \\
Gutenberg (PG-19) & 5.631 & -0.182 & [5.375, 5.888] & [-0.448, 0.085] & 0.002 & 5.450 & 5.631 \\
OpenSubtitles & 6.629 & -0.125 & [6.546, 6.712] & [-0.231, -0.018] & 0.003 & 6.504 & 6.629 \\
Wikipedia (en) & 5.401 & -0.024 & [5.317, 5.486] & [-0.144, 0.095] & 0.000 & 5.377 & 5.401 \\
DM Mathematics & 4.391 & -0.029 & [4.334, 4.447] & [-0.109, 0.051] & 0.000 & 4.361 & 4.391 \\
Ubuntu IRC & 6.371 & -0.200 & [5.943, 6.800] & [-0.633, 0.233] & 0.001 & 6.172 & 6.371 \\
BookCorpus2 & 6.300 & -0.260 & [5.789, 6.811] & [-0.778, 0.257] & 0.001 & 6.040 & 6.300 \\
EuroParl & 4.754 & -0.141 & [4.614, 4.893] & [-0.291, 0.009] & 0.003 & 4.613 & 4.754 \\
HackerNews & 7.112 & -0.074 & [7.065, 7.158] & [-0.140, -0.009] & 0.002 & 7.038 & 7.112 \\
YoutubeSubtitles & 6.444 & 0.038 & [6.324, 6.563] & [-0.101, 0.176] & 0.000 & 6.481 & 6.444 \\
PhilPapers & 6.943 & -0.097 & [6.678, 7.208] & [-0.371, 0.177] & 0.000 & 6.846 & 6.943 \\
NIH ExPorter & 6.485 & -0.039 & [6.441, 6.530] & [-0.102, 0.024] & 0.001 & 6.447 & 6.485 \\
Enron Emails & 5.819 & -0.449 & [5.698, 5.940] & [-0.621, -0.277] & 0.015 & 5.370 & 5.819 \\
\bottomrule
\end{tabular}
    \caption{Regression results on MIN20\% for the Pythia-6.9B model. $\beta_1$ is the difference in the memorization metric across train and test sets, and represents the causal effect of upweighting on the memorization of a document. The effects of upweighting a single document on memorization is generally small.}
    \label{tab:mink_regression}
\end{table}
\newpage

\subsection{Additional results on simulated ablations}
\label{appendix:curation}

\begin{table}[H]
\centering
\begin{tabular}{lcccc}
\toprule
\textbf{Threshold} & \textbf{Loss} & \textbf{MinK} & \textbf{Accuracy} & \textbf{Verbatim} \\
\midrule
50 & -0.684*** & -0.532* & 0.556** & 0.181 \\
70 & -0.725*** & -0.602** & 0.645** & 0.278 \\
90 & -0.728*** & -0.618** & 0.666*** & 0.327 \\
\bottomrule
\end{tabular}
\caption{Dataset-level correlations across different similarity score thresholds for Elasticsearch. Documents in Elasticsearch are indexed according to BM25, and the thresholds control the radius of the neighborhoods to return. The correlation increases as the threshold increases (smaller neighborhood). Significance levels: *  p < 0.05,  ** p < 0.01, *** p < 0.001.}
\label{tab:bm25_thresholds}
\end{table}

\begin{table}[h!]
\centering
\begin{tabular}{lll}
\toprule
\textbf{Metric} & \multicolumn{1}{l}{\textbf{Dataset-level}} & \multicolumn{1}{l}{\textbf{Document-level}} \\
\midrule
Loss & -0.6836*** & -0.6112*** \\
MinK\% (K=20) & -0.5322* & -0.4067*** \\
Token accuracy & 0.556** & 0.2964*** \\
Verbatim memorization & 0.1808 & 0.09846*** \\
\bottomrule
\end{tabular}
\caption{Correlations of memorization metrics with nearest neighbor counts (using BM25 > 50) at the dataset and document level. Our Elasticsearch index returns neighbors for each document, and we aggregate neighbors over documents in the same dataset to obtain dataset-level neighbor counts. Correlation on the dataset level is stronger and verbatim memorization is the weakest metric. Significance levels: *  p < 0.05,  ** p < 0.01, *** p < 0.001.}
\label{tab:correlations}
\end{table}
\begin{figure}[ht]
\centering
\includegraphics[]{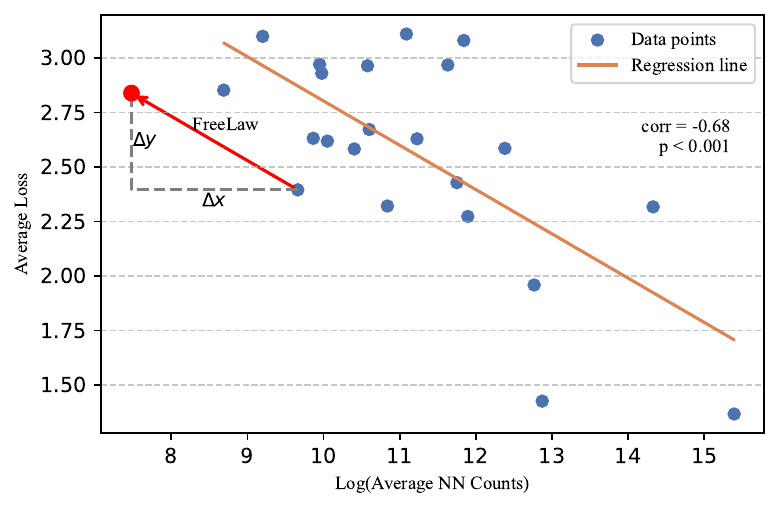}
\caption{Plot of the average model loss on a dataset against its average nearest-neighbor counts, for each of the 22 dataset components. The orange line is a best-fit regression line. To simulate an ablation, we use Table \ref{fig:dataset_overlaps} to obtain a counterfactual neighborhood count and then use the regression to calculate $\Delta Y$. The shift of the loss is parallel to the regression line. The results of each simulated ablation are presented in Figure \ref{fig:ablation_plot}.}
\label{fig:scatter}
\end{figure}

\end{document}